\documentclass[journal]{IEEEtran}
\usepackage{amsmath,amsfonts}
\usepackage{algorithmic}
\usepackage{algorithm}
\usepackage{array}
\usepackage[caption=false,font=normalsize,labelfont=sf,textfont=sf]{subfig}
\usepackage{textcomp}
\usepackage{stfloats}
\usepackage{url}
\usepackage{verbatim}
\usepackage{graphicx}
\usepackage{cite}
\usepackage{bm}
\usepackage{diagbox} 
\usepackage{multirow}
\usepackage{mathtools}
\usepackage{xcolor}

\begin{document}

\title{Matching-free Acquisition of Channels with Anisotropic Wavefronts}

\author{
    Heling Zhang,~\IEEEmembership{Student Member,~IEEE,}
    Shidong Zhou,~\IEEEmembership{Member,~IEEE.}
    \thanks{A portion of this work has been submitted to IEEE Global Communications Conference (IEEE GLOBECOM’26) for possible publication.}
    \thanks{The research presented in this paper has been kindly supported by the projects as follows, National Natural Science Foundation of China under Grants (62394294, 62394290), the Fundamental Research Funds for the Central Universities under Grant 2242022k60006.}
}

\markboth{Submitted to IEEE Transactions on Wireless Communications}%
{Zhang and Zhou \MakeLowercase{\textit{et al.}}: Matching-free Acquisition of Channels with Anisotropic Wavefronts}

\maketitle

\begin{abstract}
The escalating data rate demands of future wireless communications necessitate the deployment of extremely large aperture arrays (ELAAs) in communication systems. Acquiring accurate channel state information is crucial to execute effective precoding for such systems, in which the near-field curvature effects on the channel must be considered. Current channel estimation algorithms are generally restricted to the spherical wavefront channel (SWC), which is appropriate for isotropic scatterers, point sources, and planar reflecting surfaces. However, in practical scenarios involving curved reflecting surfaces, the reflected waves exhibit anisotropic rather than spherical wavefronts, significantly degrading the accuracy of conventional SWC-based algorithms. To tackle this challenge, we first derive a parameterized model for the anisotropic wavefront channel (AWC). Based on this model, we then propose the matching-free acquisition of channels with anisotropic wavefronts (MACAW) algorithm. Unlike conventional dictionary-based matching pursuit techniques, MACAW recovers channel parameters through fast-Fourier-transform-based frequency analysis. This approach enables precise channel estimation in AWC scenarios while maintaining a significantly lower computational complexity than existing methods. Simulation results illustrate how physical characteristics of the propagation environment influence the degree of wavefront anisotropy, and demonstrate the effectiveness of the proposed algorithm.
\end{abstract}


\section{Introduction}
To fulfill the escalating requirements of future wireless wireless systems-such as ultra-high data rates, ultra-low latency and high reliability, extremely large aperture array (ELAA) and millimeter-wave (mmWave) have emerged as pivotal technologies for the sixth-generation (6G) wireless systems. Evolving from the massive MIMO of the fifth-generation (5G) system, ELAA further scales up the number of antennas drastically. This unprecedented scale facilitates highly directional beamforming, thereby yielding substantial gains in spectral efficiency. Concurrently, the exploitation of the mmWave band unlocks vast available bandwidth, fundamentally boosting system capacity. Furthermore, the millimeter-level wavelength dictates a proportionally smaller antenna spacing, which renders the physical deployment of ELAA highly feasible within a constrained physical space \cite{Parvini25}.

In mmWave-ELAA systems, acquiring accurate channel state information is a fundamental prerequisite for precoding and beamforming. Effective channel estimation relies on exploiting two key propagation characteristics. Specifically, these channels exhibit spatial sparsity, encompassing only a few dominant propagation paths \cite{Sloane23}. Meanwhile, due to the massive physical aperture and short carrier wavelength, they are significantly affected by the wavefront curvature of each path \cite{Zhou15}. Existing algorithms are typically based on the sparse spherical wavefront channel (SWC) \cite{Liu23} model, where the wavefront curvature of each path is identical in all directions and can be represented by a single parameter. Under this model, they estimate the channel via sparse recovery techniques by exploiting its sparsity in the angle-distance (polar) domain.

However, such a spherical model is inaccurate in environments with curved reflecting surfaces. Waves reflected by such surfaces exhibit anisotropic wavefronts, meaning their curvatures vary across different directions \cite{Deschamps72}. For example, upon the incidence of a plane wave on a cylinder, the reflected cylindrical wavefront has zero curvature in one direction but non-zero curvatures in others. This wavefront anisotropy alters the phase distribution of the channel response across the antenna array, causing such anisotropic wavefront channels (AWCs) to significantly deviate from the SWC model. Consequently, the sparsity in the polar domain is destroyed, severely degrading the performance of existing methods. Specifically, since practical systems often adopt uniform planar arrays (UPAs) to achieve three-dimensional beamforming capabilities \cite{Balanis25}, the extension of the array in two directions makes the impact of wavefront anisotropy on the channel particularly pronounced. Furthermore, to save hardware overhead, systems typically employ hybrid precoding architectures \cite{Ahmed18}, which significantly reduces the number of available channel observations. Given that the channel becomes more complex due to wavefront anisotropy, channel estimation becomes substantially difficult.

In particular, when confronted with AWC, existing channel estimation frameworks exhibit two primary limitations:
\begin{itemize}
    \item \textit{Model mismatch}: To the best of our knowledge, there is currently no channel model or channel estimation method designed for AWC. Although some studies have presented beamforming and channel estimation algorithms for UPA, they predominantly assume ideal spherical wavefronts and fail to take the anisotropy of wavefront curvature into consideration. For instance, the authors in \cite{Wu23} derived an approximate near-field channel model for UPA-SWC and developed corresponding codebooks and multiple-access frameworks. \cite{Huang24, Chen25, Guo23, Peng24, Guellil26} proposed near-field channel estimation methods for UPA. These approaches inherently rely on the specific manifold structure of the SWC, and thus cannot be directly applied to AWC. Other ELAA channel estimation studies are based on the uniform linear array (ULA) model \cite{Cui22, Zhu25, Zhang24, Hussain24, Qu24, Cao23, Wang25, Xi23}. Although it is possible to apply ULA-based channel estimation to each column of the UPA in AWC scenarios, this method cannot utilize the inter-column structure of the AWC, failing to harness the coherent gain over the entire array.
    
    \item \textit{Prohibitive Complexity of Grid-Matching}: To estimate path parameters, the majority of existing near-field methods rely on dividing the polar domain into discrete grids. These algorithms then match the channel observations against these grid points to extract the path parameters. For instance, the authors in \cite{Cui22, Wang25} construct codebooks based on sampled polar domain grids and employ orthogonal matching pursuit (OMP) to iteratively identify the optimal path parameters. Similarly, the method in \cite{Xi23} applies the discrete fractional Fourier transform (DFrFT) to channel observations, recovering parameters via exhaustive 2D peak-searching over the polar grid.
    
    Inevitably, these grid-matching-based algorithms incur a computational complexity proportional to the number of grid points. To alleviate the dimensionality burden, a substantial body of literature attempts to retain grid matching solely in the angular domain while resolving the distance parameter independently. For instance, \cite{Chen25} leverages wideband channel observations to pre-determine path delays for distance computation; \cite{Zhu25, Zhang24, Cao23} treat the path distance as a learnable parameter, continuously refining it alongside angular domain matching; and \cite{Qu24} decomposes the two-dimensional matching over the angle-distance grid into two sequential one-dimensional searches. 
    
    However, these two channel estimation paradigms faces multiple challenges in AWC scenarios. First, for algorithms relying on polar-domain grids, the increased number of channel parameters in the AWC expands the grid dimensionality, resulting in prohibitive matching complexity. Second, since the wavefront curvature in the AWC is no longer determined by a single distance parameter, algorithms that resolve the distance parameter become infeasible. Furthermore, even if matching is restricted solely to the angular domain, the ultra-high angular resolution of ELAA still incurs an unacceptable computational overhead.
    
    Beyond these approaches, matching-free near-field channel estimation algorithms have also been explored. By performing phase differencing on channel observations, \cite{Huang24, Guo23, Guellil26} transform the distance and angle estimation into a harmonic analysis problem, thereby avoiding the complexity introduced by matching. However, they assume a fully digital beamforming architecture for the antenna array, and fail to generalize their methods to hybrid precoding scenarios, which are more feasible for ELAA systems.
\end{itemize}

To address these challenges, our main contributions are summarized as follows:
\begin{itemize}
    \item \textit{AWC Modeling and Wavefront Anisotropy Analysis}: We derive a wideband single-input multiple-output SIMO channel model for the AWC and reveal both theoretically and experimentally how different physical characteristics of the propagation scenario affect wavefront anisotropy. This analysis establishes a criterion to determine the applicable scope of the conventional SWC model. Specifically, we first formulate the single-path single-bounce channel, which is subsequently generalized to a generic multipath multiple-bounce AWC model. Following this, we derive a theoretical lower bound for the maximum cosine similarity between the steering vectors of the AWC and the SWC. The analytical expression of this bound explicitly captures the mechanism by which various physical characteristics influence the degree of wavefront anisotropy. Finally, simulation results validate the correctness of this theoretical bound, and corroborate our analytical conclusions regarding the relationship between these physical characteristics and wavefront anisotropy.
    
    \item \textit{Low-Complexity Estimation Algorithm for AWC}: We propose the matching-free acquisition of channels with anisotropic wavefronts (MACAW) algorithm, which is tailored for the AWC and features remarkably low complexity. Specifically, the algorithm exploits wideband information to separate distinct propagation paths and obtains their delay estimates. Subsequently, the wavefront curvature matrix and arrival direction for each path are extracted sequentially. By employing phase differencing, the extraction of the curvature matrix is simplified to a two-dimensional (2D) spectral peak search; furthermore, based on the estimated curvature matrix, the near-field spectral broadening can be eliminated, enabling the AoAs to also be obtained via spectral peak search. Since these peak searches can be efficiently implemented using the Fast Fourier Transform (FFT), the computational complexity is significantly reduced. All parameters are then refined through our specifically designed low-complexity optimizer, simultaneously acquiring the path gains and recovering the complete channel. Simulation results demonstrate that in AWC scenarios, the estimation performance of the proposed algorithm approaches the Cramér-Rao lower bound (CRLB) when the signal-to-noise ratio (SNR) is $\ge -5$ dB.
\end{itemize}

\textit{Organization}: The remainder of this paper is organized as follows. Section \ref{sec:AWC} formulates the AWC models based on the geometrical optics approximation, and quantifies the modeling discrepancy between AWC and traditional SWC. Section \ref{sec:MACAW} details the proposed matching-free channel estimation algorithm. Section \ref{sec:Simulation} presents simulation results to validate the effectiveness and superiority of the proposed method. Finally, Section \ref{sec:Conclusion} concludes the paper.

\textit{Notation}: Boldface lower-case and upper-case letters denote column vectors and matrices, respectively. For a matrix $\mathbf{A}$, $\mathbf{A}^T$, $\mathbf{A}^H$ and $\mathbf{A}^{-1}$, denote its transpose, conjugate transpose and inverse, respectively. $[\mathbf{A}]_{i,j}$ represents the element in the $i$-th row and $j$-th column of $\mathbf{A}$, and $[\mathbf{A}]_{r_1{:}r_2, c_1{:}c_2}$ denotes the sub-matrix of $\mathbf{A}$ consisting of elements from rows $r_1$ to $r_2$ and columns $c_1$ to $c_2$. $\Vert\mathbf{A}\Vert_F$ denotes the Frobenius norm of $\mathbf{A}$. The operation $\mathbf{A}\odot\mathbf{B}$ denotes the Hadamard product. $\text{tr}(\mathbf{A})$ denotes the trace of $\mathbf{A}$. $\text{vec}(\mathbf{A})$ denotes the vectorization of matrix $\mathbf{A}$ by stacking its columns, and $\text{unvec}(\cdot)$ denotes the corresponding inverse operation. For a vector $\mathbf{a}$, $\|\mathbf{a}\|_2$ denotes its Euclidean norm, and $\text{diag}(\mathbf{a})$ represents a diagonal matrix with the elements of $\mathbf{a}$ on its main diagonal. $\mathbf{I}_N$ is the $N \times N$ identity matrix. $J_n(\cdot)$ denotes the $n$-th order Bessel function of the first kind. $c_0$ denotes the speed of light.

\begin{figure}[!t]
\centering
\includegraphics[width=\linewidth]{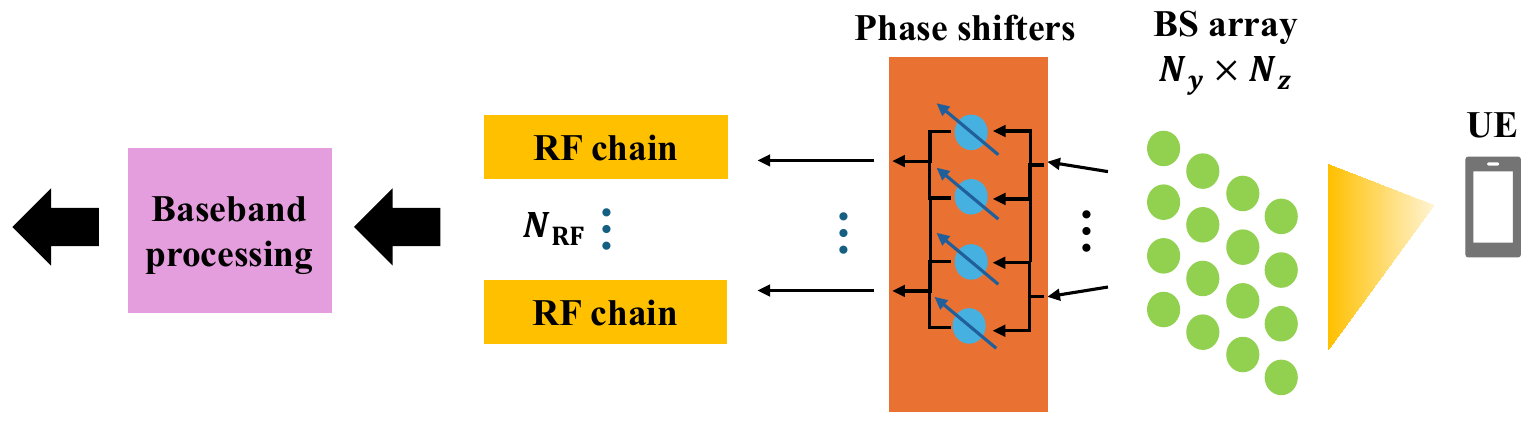}
\caption{ELAA system with hybrid precoding.}
\label{img:Hybrid_precoding}
\end{figure}

\section{System Model}\label{sec: System Model}
We consider an orthogonal frequency division multiplexing (OFDM) communication system with ELAA and a hybrid precoding architecture as Fig. \ref{img:Hybrid_precoding}. The base station (BS) is equipped with a UPA having $N_y\times N_z= N$ antennas, which are connected to $N_{\text{RF}}$ RF chains via an analog phase-shifting network. The antenna elements of the UPA are separated by a distance of $d_\text{ant}$ along both dimensions, while the user equipment (UE) has a single antenna. Both the BS and the UE employ isotropic antennas. The total OFDM bandwidth is $B$, and $M$ pilot-carrying subcarriers are uniformly distributed with a spacing of $\Delta f$, centered at $f_c$. $B$ is substantially smaller than the center frequency $f_c$, thereby limiting the impact of the beam split effect \cite{Dai22}. To estimate the channel, the UE transmits unit pilots over $P$ consecutive OFDM symbols. The channel $\mathbf{h}_m$ on the $m$-th carrier is the aggregation of sub-channels from $K_\text{path}$ paths, and is formulated in Section \ref{sec:AWC}.

On the $m$-th subcarrier of the $p$-th OFDM symbol, the received signal at the BS is given by
\begin{equation}
    \mathbf{y}_{p,m} = \mathbf{W}_p(\mathbf{h}_m +\mathbf{n}_{p,m}),
\end{equation}
where $\mathbf{n}_{p,m}$ is the additive white Gaussian noise with variance $\sigma^2_n$. $\mathbf{W}_p$ is the hybrid precoding matrix at the $p$-th OFDM symbol, whose entries are constant-modulus. In our system, $\mathbf{W}_p$ is designed to be a semi-orthogonal matrix, which will be detailed in the end of this section. Hence, the noise after hybrid precoding $\tilde{\mathbf{n}}_{p,m}=\mathbf{W}_p\mathbf{n}_{p,m}$ is also white Gaussian. 

By stacking all received signals over $P$ consecutive OFDM symbols together, we obtain the aggregated narrowband channel observation model:
\begin{equation}
    \mathbf{y}_{m} = \mathbf{W}\mathbf{h}_m+\tilde{\mathbf{n}}_{m},
\end{equation}
where $\mathbf{y}_{m} = [\mathbf{y}_{1,m}^T,\dots,\mathbf{y}_{p,m}^T]^T$, $\mathbf{W} = [\mathbf{W}_{1}^T,\dots,\mathbf{W}_{p}^T]^T$ and $\tilde{\mathbf{n}}_{m} = [\tilde{\mathbf{n}}_{1,m}^T,\dots,\tilde{\mathbf{n}}_{p,m}^T]^T$. 
the narrowband observations over $M$ subcarriers are further concatenated to obtain the wideband channel observation:
\begin{equation}
    \mathbf{Y} =\mathbf{WH}+\tilde{\mathbf{N}},
\end{equation}
where $\mathbf{Y} = [\mathbf{y}_{1},\dots, \mathbf{y}_{m}]$, $\mathbf{H} = [\mathbf{h}_1,\dots, \mathbf{h}_m]$ and $\tilde{\mathbf{N}}=[\tilde{\mathbf{n}}_{1},\dots,\tilde{\mathbf{n}}_{m}]$. Because the noise is independent across different OFDM symbols and subcarriers, each element of $\tilde{\mathbf{N}}$ is an independent and identically distributed (i.i.d.) Gaussian random variable with variance $\sigma_n^2$.

We also make two realistic assumptions regarding the paths:
\begin{itemize}
    \item The path length difference between any two paths is at least equal to the distance resolution of the system provided by $\frac{c_0}{B}$. This ensures that each path can be separated utilizing the frequency domain information. For a system with $B = 100\text{ MHz}$, this requires a minimum path length difference of $3\text{ m}$.
    \item All scatterers in the propagation scenario have convex reflection surfaces (e.g. cylinders or spheres), and the distances from all reflection points to the BS array center are greater than $r_{\min}$. This assumption restricts the possible reflection wavefront to be convex with limited positive curvatures. In our algorithm, $r_{\min}$ is set to $\frac{1}{5}r_{\text{nf}}$, where $r_{\text{nf}} = \frac{1}{2}\sqrt{D^3/\lambda}$ is the minimum propagation distance for a spherical wavefront to be accurately approximated by an isotropic paraboloid on a UPA with a diagonal length of $D$ \cite{Selvan17}. For a $1\text{ m} \times 1\text{ m}$ UPA and a carrier frequency of $30\text{ GHz}$, $r_{\min}$ is calculated as $1.7\text{ m}$.
\end{itemize}

Specifically, since the combining matrix $\mathbf{W}$ can be flexibly designed, it is configured as a subsampled randomized Fourier transform (SRFT) matrix. Such a matrix is a fast Johnson-Lindenstrauss transform, which offers the following properties:
\begin{itemize}
\item It satisfies the following condition with high probability:
    \begin{equation}
        \vert\mathbf{a}_1^H\mathbf{W}^H\mathbf{W}\mathbf{a}_2\vert\leq \epsilon,
    \end{equation}
for any two orthonormal basis vectors $\mathbf{a}_1$ and $\mathbf{a}_2$ chosen from a certain basis set $\mathcal{A}$ \cite{Woodruff14}. Since $\epsilon$ is a small positive constant, this inequality implies that orthogonality is well-preserved after subsampling. In particular, by setting $\mathbf{a}_1$ and $\mathbf{a}_2$ to be orthogonal Fourier basis vectors, it can be deduced that the spectrum of $\mathbf{W}^H\mathbf{y}$ closely preserves the spectral features of $\mathbf{h}$.
\item $\mathbf{W}^H\mathbf{y}$ can be efficiently computed via FFT. 
\item Each sub-matrix $\mathbf{W}_p$ is semi-orthogonal, which ensures that the noise remains white after the hybrid precoding stage, thereby eliminating the need for pre-whitening.
\end{itemize}

\section{AWC Formulation} \label{sec:AWC}
In this section, we extend the electromagnetic computation framework proposed in Section III of \cite{Deschamps72} to ELAA systems to model the channel vector $\mathbf{h}_m$ of AWC introduced in Section \ref{sec: System Model}. Utilizing the distance-curvature relationship provided in \cite{Deschamps72}, we first formulate the single-path channel and derive the AWC steering vector for a given wavefront curvature at the BS center. Subsequently, for modeling completeness, we follow the principles in \cite{Deschamps72} to briefly describe the geometric derivation of this curvature from the environment. Finally, the impact of various physical characteristics on the degree of wavefront anisotropy is theoretically investigated.

\textit{Remark 1}: The derived AWC model relies on two prerequisites: 1) the GO approximation holds, and 2) the incident wavefront for each reflection surface and the reflection surface itself can both be approximated as paraboloids. We assume these prerequisites are satisfied throughout our analysis, and scenarios violating these approximations fall beyond the scope of this paper. 

\textit{Preliminary}: The anisotropic curvature of a convex wavefront or a reflecting surface is described by a positive definite curvature matrix $\mathbf{Q}$, which corresponds to the paraboloidal approximation of the local wavefront. Specifically, a wavefront passing through a given point $O$ can be locally approximated as a paraboloid with its vertex at $O$. Let us establish a local coordinate system originating at $O$, defining the wave propagation direction $\mathbf{k}$ at this point as the longitudinal axis $h$. By specifying two orthogonal basis vectors, $\mathbf{u}_1$ and $\mathbf{u}_2$, on the plane perpendicular to the propagation direction, the approximated paraboloidal surface can be expressed as $h = -\frac{1}{2}\mathbf{u}^T\mathbf{Q}\mathbf{u}$. Here, $\mathbf{u} \in \mathbb{R}^2$ represents the projection coordinates of a point on the surface with respect to the bases $\mathbf{u}_1$ and $\mathbf{u}_2$. The entries of the curvature matrix $\mathbf{Q}$ depend on the choice of the local coordinate basis. If $\mathbf{u}_1$ and $\mathbf{u}_2$ are selected such that $\mathbf{Q}$ is diagonalized, these basis directions are termed the principal directions of the surface, with the corresponding diagonal entries representing the principal curvatures. The principal radii of curvature, denoted as $R_1$ and $R_2$, are given by the reciprocals of these principal curvatures. In particular, when the two principal radii of a wavefront are equal, the wavefront is isotropic.

We begin from deriving the single-path channel and the AWC steering vector on a certain sub-carrier with wavelength $\lambda$. Consider a single-path channel, in which the electromagnetic wave arrives at the center of the BS array, which is denoted as $O_\text{BS}$. Let $\hat{\mathbf{k}}_{\text{BS}}$ and $\mathbf{Q}_{\text{BS}}$ denote the propagation direction and curvature matrix of the wave at $O_\text{BS}$, respectively, with $\hat{\mathbf{u}}_{\text{BS}1}$ and $\hat{\mathbf{u}}_{\text{BS}2}$ representing the coordinate basis vectors associated with the curvature matrix. In this scenario, the amplitude of the electric field across the BS array can be regarded as uniform, and the phase at each antenna is determined by the path distance. Therefore, the channel is formulated as:
\begin{equation}\label{eq: phase-eikonal}
[\tilde{\mathbf{C}}]_{n_y,n_z} = \tilde{a} e^{-j\frac{2\pi}{\lambda}S(\mathbf{p}_{n_y, n_z})},
\end{equation}
where $\tilde{\mathbf{C}}$ is the single-path channel matrix, and its element with subscript $n_y, n_z$ represents the channel response at the corresponding antenna element. $\mathbf{p}_{n_y, n_z}$ denotes the position coordinates of this antenna relative to the array center. The path amplitude $\tilde{a}$ depends on the shapes, electromagnetic properties, and polarization configurations of the reflecting surfaces along the propagation path \cite{Balanis12}. Furthermore, the distance function $S(\mathbf{p})$ can be approximated via Taylor expansion in the neighborhood of $O_\text{BS}$, where its gradient and Hessian are given by the wavefront normal and the curvature matrix \cite{Deschamps72}:
\begin{equation}\label{eq:eikonal expansion} 
    S(\mathbf{p}) \approx S(\mathbf{0}) + \hat{\mathbf{k}}_{\text{BS}}^T \mathbf{p} + \frac{1}{2} \mathbf{p}^T (\mathbf{U}_\text{BS}\mathbf{Q}_\text{BS}\mathbf{U}_\text{BS}^T)\mathbf{p}
\end{equation}
where $\mathbf{U}_\text{BS}=[\hat{\mathbf{u}}_{\text{BS}1},\hat{\mathbf{u}}_{\text{BS}2}]$. The reference distance $S(\mathbf{0})$ corresponds to the total propagation distance from the UE to the BS array center.

Let $\hat{\mathbf{y}}$ and $\hat{\mathbf{z}}$ denote the extension directions of the BS array. For UPA, the position of the $(n_y, n_z)$-th antenna element relative to the array center is given by $\mathbf{p}_{n_y,n_z} = d_{\text{ant}}( \delta_{n_y} \hat{\mathbf{y}} + \delta_{n_z} \hat{\mathbf{z}} )$, where $\delta_{n_y} = n_y -\frac{N_y+1}{2}$ and $\delta_{n_z} = n_z -\frac{N_z+1}{2}$. Consequently, combining (\ref{eq: phase-eikonal}) and (\ref{eq:eikonal expansion}) , we obtain the following normalized channel determined by the effective direction and effective curvature, which characterizes the phase distribution of the channel response across the array:
\begin{equation}\label{eq: effective AWC steering matrix}
[\mathbf{C}]_{n_y, n_z} = \frac{1}{\sqrt{N}}e^{-j 2\pi \left( \bar{\mathbf{k}}^T \mathbf{n}_{y,z} + \frac{1}{2} \mathbf{n}_{y,z}^T \bar{\mathbf{Q}} \mathbf{n}_{y,z} \right)}.
\end{equation}
where $\mathbf{C}$ is the normalized version of $\tilde{\mathbf{C}}$, and $\mathbf{n}_{y,z} = [\delta_{n_y},\delta_{n_z}]^T$ defines the two-dimensional discrete index vector. The effective direction vector, given by
\begin{equation} \label{eq:wavenumber}
    \bar{\mathbf{k}} = \frac{d_{\text{ant}}}{\lambda}\begin{bmatrix} \hat{\mathbf{k}}_{\text{BS}}^T \hat{\mathbf{y}} \\ \hat{\mathbf{k}}_{\text{BS}}^T \hat{\mathbf{z}} \end{bmatrix}
\end{equation}
and the effective curvature matrix, given by
\begin{equation} \label{eq:curvature}
    \bar{\mathbf{Q}} = \frac{d_{\text{ant}}^2}{\lambda}
    \mathbf{P}^T\mathbf{Q}_{\text{BS}}\mathbf{P} \  \left(\mathbf{P}= 
\mathbf{U}_\text{BS}^T\left[ \hat{\mathbf{y}},\hat{\mathbf{z}}\right]
\right)
\end{equation}
correspond to the relative angle of the incident electromagnetic wave arriving at the array and its effective curvature across the array aperture, respectively. 

By vectorizing $\mathbf{C}$, we obtain the AWC steering vector $\mathbf{c} = \mathrm{vec}(\mathbf{C})$. To explicitly highlight its dependence on the direction and curvature parameters, we hereafter denote this steering vector as
\begin{equation}
    \mathbf{c} = \mathbf{c}(\bar{\mathbf{k}}, \bar{\mathbf{Q}}).
\end{equation}
Accordingly, the single-path channel can be expressed as $a\mathbf{c}(\bar{\mathbf{k}}, \bar{\mathbf{Q}})$, where $a = \sqrt{N}e^{-j2\pi S(\mathbf{0})/\lambda}\tilde{a}$ denotes the effective complex path gain.

\begin{figure}[!t]
\centering
\includegraphics[width=3.0in]{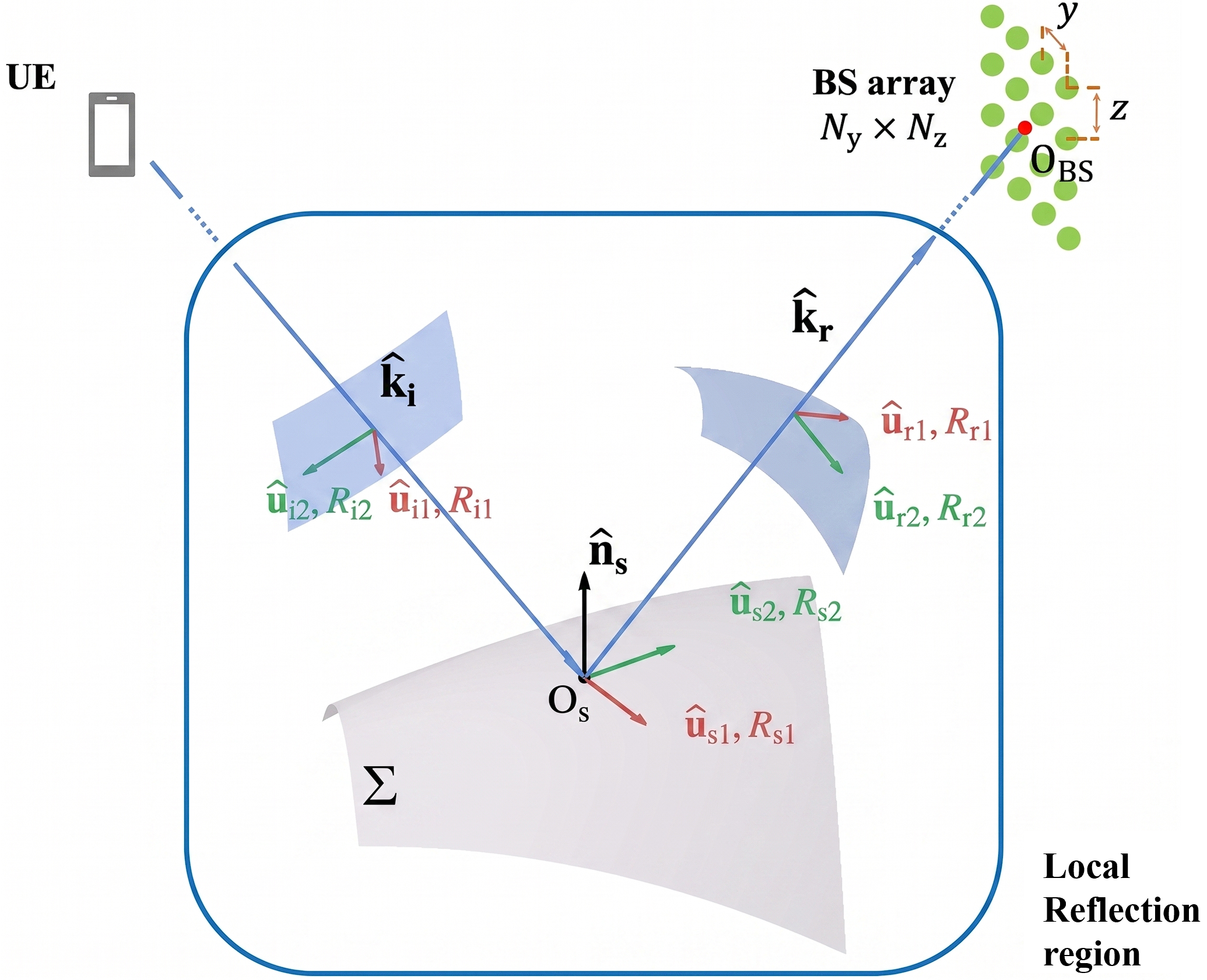}
\caption{Reflection on a curved surface.}
\label{img:curved_reflection}
\end{figure}
\begin{figure}[!t]
\centering
\includegraphics[width=2.5in]{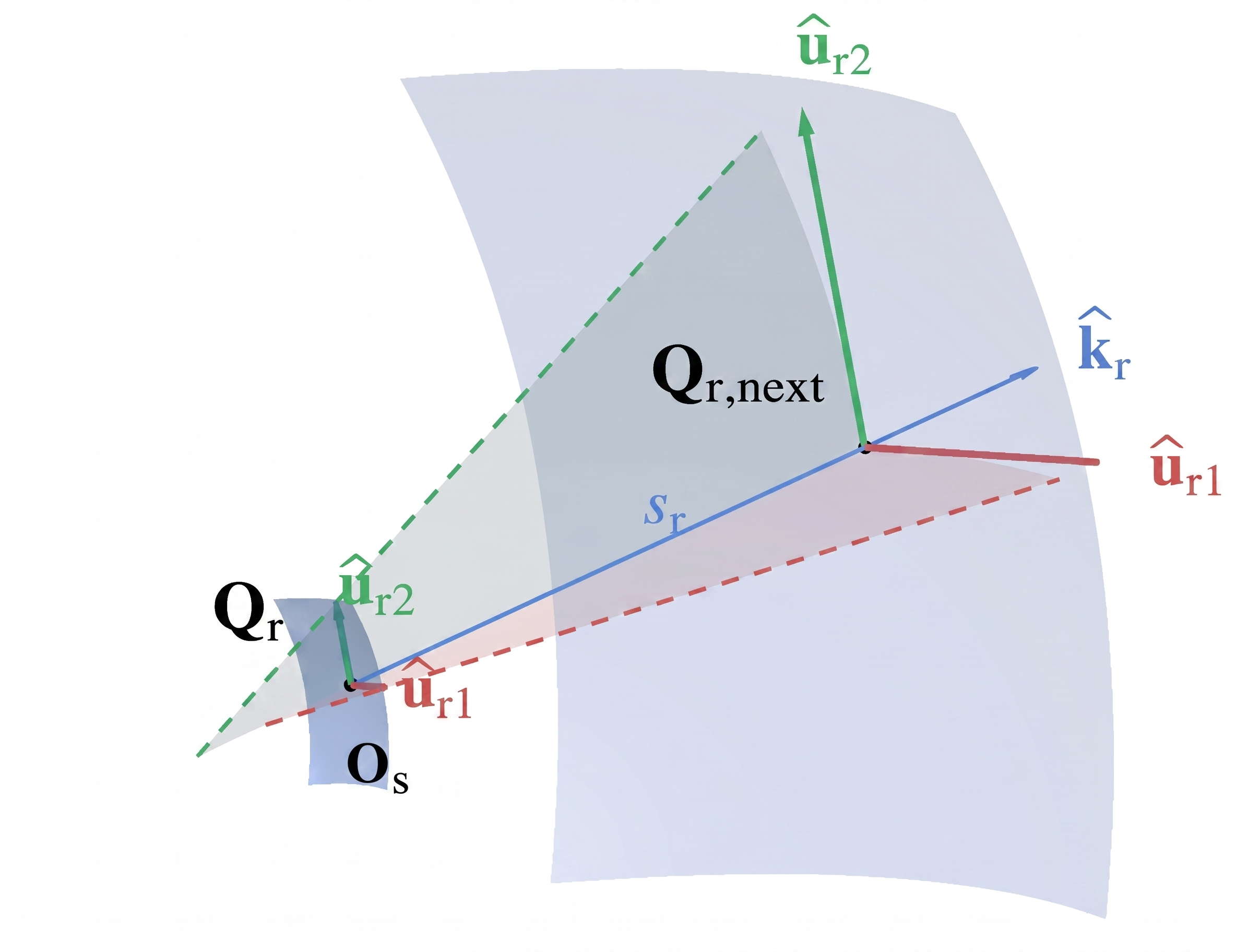}
\caption{The evolution of the wavefront during propagation.}
\label{img:caustic_propagation}
\end{figure}

The electromagnetic wave arriving at the BS typically undergoes multiple reflections during propagation, which dictate its arrival direction and wavefront curvature. To determine $\mathbf{Q}_{\text{BS}}$, we derive the multiple-bounce channel utilizing the method proposed in \cite{Deschamps72}. The overall propagation process can be decomposed into an alternation of two distinct phases: 
\begin{itemize}
    \item \textit{Reflection}, where the electromagnetic wave impinges upon and is reflected by a curved surface as in Fig. \ref{img:curved_reflection};
    \item \textit{Propagation}, where the reflected wave travels to the subsequent reflecting surface as in Fig. \ref{img:caustic_propagation};
\end{itemize}

To establish the general multiple-bounce channel model, we first investigate the evolution of the curvature matrix during the reflection phase. In this phase, an incident electromagnetic wave characterized by a curvature matrix $\mathbf{Q}_{\text{i}}$ strikes a convex surface characterized by a curvature matrix $\mathbf{Q}_{\text{s}}$. The curvature matrix $\mathbf{Q}_{\text{r}}$ of the reflected wavefront is given by:
\begin{equation} \label{eq:Deschamps}
   \mathbf{Q}_{\text{r}} = \mathbf{Q}_{\text{i}} + 2(\mathbf{\Theta}^{-1})^T \mathbf{Q}_{\text{s}} \mathbf{\Theta}^{-1} \cos \theta_{\text{i}},
\end{equation}
where $\theta_{\text{i}}$ denotes the angle of incidence on the convex surface. The transformation matrix $\mathbf{\Theta}$ is defined as:
\begin{equation}
\mathbf{\Theta} = \begin{bmatrix}
\hat{\mathbf{u}}_{\text{i}1}^T \hat{\mathbf{u}}_{\text{s}1} &
\hat{\mathbf{u}}_{\text{i}1}^T \hat{\mathbf{u}}_{\text{s}2} \\
\hat{\mathbf{u}}_{\text{i}2}^T \hat{\mathbf{u}}_{\text{s}1} &
\hat{\mathbf{u}}_{\text{i}2}^T \hat{\mathbf{u}}_{\text{s}2} \end{bmatrix}.
\end{equation}
Here, $\hat{\mathbf{u}}_{\text{i}m}$ and $\hat{\mathbf{u}}_{\text{s}m}$ ($m \in \{1, 2\}$) represent the chosen local coordinate basis vectors for the incident wavefront and the reflecting surface, respectively. The local coordinate basis for the reflected wavefront, corresponding to $\mathbf{Q}_{\text{r}}$, is determined by the standard reflection law:
\begin{equation}
\hat{\mathbf{u}}_{\text{r}m} = \hat{\mathbf{u}}_{\text{i}m} - 2(\hat{\mathbf{n}}_{\text{s}}^T \hat{\mathbf{u}}_{\text{i}m})\hat{\mathbf{n}}_{\text{s}}, \quad m \in \{1, 2\},
\end{equation}
where $\hat{\mathbf{n}}_{\text{s}}$ is the unit normal vector of the reflecting surface at the point of incidence. 

Subsequently, we examine the evolution of the curvature matrix during the propagation phase. As the reflected electromagnetic wave travels a distance $s_{\text{r}}$ to the next reflecting surface (or the BS center), its two principal radii of curvature uniformly increase by $s_{\text{r}}$, while the principal directions remain unchanged. This transformation is described by:
\begin{equation}\label{eq:Deschamps2}
   \mathbf{Q}_{\text{r,next}} = (\mathbf{Q}_{\text{r}}^{-1} + s_{\text{r}} \mathbf{I}_2)^{-1}.
\end{equation}

Based on the aforementioned discussions, the calculation procedure for $\mathbf{Q}_{\text{BS}}$ can be summarized as follows:
\begin{itemize}
    \item For the line-of-sight (LoS) path, the electromagnetic wave incident on the array is a conventional spherical wave. The radius of the sphere is exactly the distance between the BS and the UE, denoted by $R_{\text{LoS}}$. Consequently, the curvature matrix is given by $\mathbf{Q}_{\text{BS}} = \frac{1}{R_{\text{LoS}}} \mathbf{I}_2$.
    \item For the multiple-bounce reflected paths, the incident wave at the first reflection is a spherical wave originating from the UE, with its initial radius being the distance to the first reflecting surface (denoted as $s_1$). Thus, the initial curvature matrix is formulated as $\mathbf{Q}_0 = \frac{1}{s_1} \mathbf{I}_2$. Subsequently, the evolution of the curvature matrix across multiple bounces is calculated by iteratively applying (\ref{eq:Deschamps}) and (\ref{eq:Deschamps2}), yielding the final $\mathbf{Q}_{\text{BS}}$ at the BS array.
\end{itemize}
Since the aim of this paper is to model the impact of wavefront curvature on channel characteristics, the derivations for the arrival directions and path gains are omitted here.

Finally, the single-path sub-channels are summed up to formulate the multipath AWC. The multipath AWC response vector $\mathbf{h}$ takes the form
\begin{equation}\label{eq:AWC channel}
    \mathbf{h} = \sum_{k=1}^{K_\text{path}} a_k \mathbf{c}(\bar{\mathbf{k}}_k, \bar{\mathbf{Q}}_k).
\end{equation}

\textit{Remark 2}: The SWC model is a special case of the AWC. Specifically, if the two principal curvatures of $\mathbf{Q}_{\text{BS}}$ are identical such that $\mathbf{Q}_{\text{BS}}$ is a scaled identity matrix, the electromagnetic wave impinging on the BS array becomes spherical, causing the AWC to degrade to the SWC. Such a condition corresponds to specific propagation scenarios: the LOS path, reflections from perfectly planar surfaces (where the surface curvature matrix is zero), or scattering from ideal point scatterers (where the surface curvature approaches infinity). Beyond these scenarios, the SWC model loses its accuracy.

At the end of this section, we discuss how various physical characteristics affect the degree of wavefront anisotropy of the channel. For a single AWC steering vector $\mathbf{c}$, we quantify the degree of wavefront anisotropy using its maximum cosine similarity with the SWC model, defined as:
\begin{equation}
    \beta = \max_{\mathbf{b}} \left| \mathbf{b}^H \mathbf{c} \right|,
\end{equation}
where $\mathbf{b}$ denotes a unit length SWC steering vector. This maximum cosine similarity directly corresponds to the beamforming energy loss incurred when the SWC model is applied to this AWC path. Consequently, it can serve as a criterion to evaluate when the AWC model needs to be adopted.

As derived in the Appendix, provided that the incident angle $\theta$ is not excessively large, (i.e., excluding grazing incidence on the BS array), the lower bound of $\beta$ can be estimated by:
\begin{equation}\label{eq:cossim expression}
\beta \geq \frac{\pi}{4}t(\mu^*)\sqrt{J_0^2(\mu^*) + J_1^2(\mu^*)},
\end{equation}
where $t(\mu^*)$ is a scaling factor, and $\mu^*$ is the anisotropic parameter defined as:
\begin{equation}
\mu^* = \frac{\pi d_{\text{ant}}^2(N_y^2+N_z^2)}{16\lambda}|q_1 - q_2|,
\end{equation}
where $q_1$ and $q_2$ are the principal curvatures of the anisotropic wavefront corresponding to $\mathbf{c}$. Based on the expression for $\mu$, the degree of wavefront anisotropy is significantly influenced by the following physical characteristics:
\begin{itemize}
    \item The antenna diagonal length $D = d_{\text{ant}} \sqrt{N_y^2+N_z^2}$, and the carrier wavelength.
    \item The difference between the principal curvatures of the incident wavefront. As discussed in Section \ref{sec:MACAW}, it is related to the curvature of the reflecting surfaces and the propagation distance.
\end{itemize}

$\beta$ serves as a practical criterion to determine whether the conventional SWC model is whether to rely on the conventional SWC model or to adopt the proposed AWC model. As an example, we evaluate the applicable range of the SWC model for a specific system. Assuming a cosine similarity tolerance of $\beta = 0.9$, (\ref{eq:cossim expression}) yields a corresponding threshold of $\mu^* = 0.59$. For a $1\,\text{m} \times 1\,\text{m}$ UPA operating at $30$ GHz, maintaining SWC accuracy requires the principal curvature difference to satisfy $|q_1 - q_2| \leq 0.0076\,\text{m}^{-1}$. According to (\ref{eq:Deschamps2}), we have $|q_1 - q_2| \leq 1/s$, where $s$ denotes the distance from the last-bounce scatterer to the BS center. Consequently, to guarantee the validity of the SWC model for arbitrary scatterers, this last-bounce distance must be at least $131.6$ m. This distance requirement can be reduced if specific physical characteristics, such as scatterer curvatures and inter-bounce propagation distances, are available.

\section{Proposed Channel Estimation Algorithm} \label{sec:MACAW}

\textit{Notation}: In this section, the vector $\mathbf{h}_{k,m}$ denotes the sub-channel of the $k$-th path on the $m$-th subcarrier. $\mathbf{H}_k= [\mathbf{h}_{1,m},\dots, \mathbf{h}_{K_\text{path},m}]$ denotes the $k$-th single-path wideband channel. $\mathbf{H}_{k,m}^{\text{unvec}} = \text{unvec}(\mathbf{h}_{k,m})$ denotes the unvectorized single-path channel.

Each path component can be expressed in the following unified form according to (\ref{eq: effective AWC steering matrix}):
\begin{equation}\label{eq:general path expression}
    [\mathbf{H}^\text{unvec}_{k,m}]_{n_y,n_z} = a_k e^{-j2\pi(1+\frac{\Delta f}{f_c} \delta_m)\left(\bar{S}_k+\bar{\mathbf{k}}^T_k \mathbf{n}_{y,z} + \frac{1}{2} \mathbf{n}_{y,z}^T \bar{\mathbf{Q}}_k \mathbf{n}_{y,z} \right)}
\end{equation}
where $\delta_m = m-\frac{M+1}{2}, m \in \left\{1,2,  \dots, M\right\}$, and $\bar{S}_k = \frac{f_c}{c_0}{S}_k(\mathbf{0})$ is the effective distance of the $k$-th path. To estimate the overall channel, it suffices to recover all $\mathbf{h}_{k,m}$ for $K_\text{path}$ paths by estimating their corresponding spatial parameters $\bar{\mathbf{k}}_k$ and $\bar{\mathbf{Q}}_k$, distance parameters $\bar{S}_k$ and complex coefficients $a_k$.

The estimation of these parameters is performed in a multi-step procedure. First, we concurrently extract the effective distance of each path from the wideband channel observations by exploiting frequency-domain information and coarsely recover its narrowband response at the center frequency, denoted as $\mathbf{h}_{k,\text{ctr}}$ ($\delta_m = 0$), to facilitate the estimation of spatial parameters. Next, for each individual path, the effective direction and curvature are estimated based on the phase differences across $\mathbf{h}_{k,\text{ctr}}$. Finally, all the obtained parameters are further refined by a custom-designed two-stage optimizer with low computational complexity, and are subsequently utilized to reconstruct the overall channel. 

The detailed procedures for each step, along with an analysis of the computational complexity, are elaborated below.

\subsection{Distance Estimation and Recovery of Single-Path Channels}
By omitting the beam split item $\frac{\Delta f}{f_c} \delta_m(\bar{\mathbf{k}}^T_k \mathbf{n}_{y,z} + \frac{1}{2} \mathbf{n}^T_{y,z} \bar{\mathbf{Q}}_k \mathbf{n}_{y,z})$, the wideband channel corresponding to the $k$-th path can be approximated as:
\begin{equation}\label{eq:wideband channel expression}
\mathbf{H}_k = \mathbf{h}_{k,\text{ctr}} \bm{\gamma}^T_k,
\end{equation}
where $\bm{\gamma}_k$ is the frequency domain phase shift vector for the $k$-th path, with its $m$-th element given by $[\bm{\gamma}_k]_m = e^{-j 2\pi\frac{\Delta f}{f_c} \delta_m \bar{S}_k}$. Since the total channel $\mathbf{H}$ is the superposition of all paths, the observation signal $\mathbf{Y}$ can be factored into the product of the compressed spatial response matrix $\mathbf{W}\mathbf{H}_{\text{spc}}$ and the frequency response matrix $\mathbf{\Gamma}$:
\begin{equation}
\mathbf{Y} = \mathbf{W}\mathbf{H}_{\text{spc}}\mathbf{\Gamma}^T + \tilde{\mathbf{N}},
\end{equation}
where $\mathbf{H}_{\text{spc}} = [\mathbf{h}_{1,\text{ctr}}, \dots, \mathbf{h}_{K_\text{path},\text{ctr}}]$ and $\mathbf{\Gamma} = [\bm{\gamma}_1, \dots, \bm{\gamma}_{K_\text{path}}]$.

The transpose of $\mathbf{Y}$ can be viewed as $P N_{\text{RF}}$ independent observations (each column being one observation) of a signal composed of $K_\text{path}$ superimposed non-coherent harmonic components. Therefore, applying the RELAX algorithm \cite{Li95} to $\mathbf{Y}^T$ can effectively separate all paths, yielding their spatial observation signals $\mathbf{Wh}_{1,\text{ctr}}, \dots, \mathbf{Wh}_{K_\text{path},\text{ctr}}$ and distance parameters $\hat{\bar{S}}_1, ..., \hat{\bar{S}}_{K_\text{path}}$.

The spatial observation of the $k$-th path $\mathbf{Wh}_{k,\text{ctr}}$, denoted as $\hat{\mathbf{y}}_{k,\text{ctr}}$, represents the compressed spatial observation of the $k$-th path. It can be modeled as:
\begin{equation}\label{eq:center freq observation}
    \hat{\mathbf{y}}_{k,\text{ctr}} = \mathbf{W}\mathbf{h}_{k,\text{ctr}} + \mathbf{n}_k,
\end{equation}
With a slight abuse of notation, $\mathbf{n}_k$ encompasses the original noise in $\mathbf{Y}$ and the estimation error introduced by RELAX. 

To facilitate subsequent estimation, we now perform a coarse recovery of $\mathbf{h}_{k,\text{ctr}}$ from $\hat{\mathbf{y}}_{k,\text{ctr}}$. Given the structure of $\mathbf{h}_{k,m}$ in (\ref{eq:general path expression}) , its energy is highly concentrated in a frequency range around the 2D frequencies corresponding to $\bar{\mathbf{k}}_k$, while the spectral spread is dictated by $\bar{\mathbf{Q}}_k$. In other words, the dominant energy of $\mathbf{h}_{k,m}$ lies within a specific subspace $\mathcal{U}_k$ spanned by a group of Fourier bases. Consequently, $\hat{\mathbf{y}}_{k,\text{ctr}}$ can be rewritten as:
\begin{equation}
    \hat{\mathbf{y}}_{k,\text{ctr}} \approx \mathbf{W}\mathbf{U}_k \bar{\mathbf{h}}_{k,\text{ctr}} + \mathbf{n}_k,
\end{equation}
where $\mathbf{U}_k$ is a Fourier column-orthogonal matrix whose columns span $\mathcal{U}_k$. Due to the orthogonality-preserving property of $\mathbf{W}$, $\mathbf{W}\mathbf{U}_k$ is approximately column-orthogonal, implying its pseudo-inverse can be approximated by $(\mathbf{W}\mathbf{U}_k)^H$. Thus, the least squares (LS) estimate of $\mathbf{h}_k$ is given by:
\begin{equation}
    \hat{\mathbf{h}}_{k,\text{ctr}} \approx \mathbf{U}_k \mathbf{U}_k^H \mathbf{W}^H \hat{\mathbf{y}}_{k,\text{ctr}}.
\end{equation}
This indicates that to obtain the optimal spatial response estimate for the $k$-th path, we should:
\begin{itemize}
    \item Decompress the observation $\hat{\mathbf{y}}_{k,\text{ctr}}$ to obtain a coarse estimate: $\hat{\mathbf{h}}_{k, \text{coarse}} = \mathbf{W}^H \hat{\mathbf{y}}_{k,\text{ctr}}$.
    \item Project $\hat{\mathbf{h}}_{k, \text{coarse}}$ onto the subspace where the energy of $\mathbf{h}_{k,\text{ctr}}$ is most concentrated.
\end{itemize}
Due to the spectrum-preserving property of $\mathbf{W}^H\mathbf{W}$, the energy concentration region in the spectrum of $\hat{\mathbf{h}}_{k, \text{coarse}}$ matches that of $\mathbf{h}_{k,\text{ctr}}$. Finding the optimal subspace and performing the projection is equivalent to locating the most energy-concentrated region in the spectrum of $\hat{\mathbf{h}}_{k, \text{coarse}}$, preserving the signal components within this region, and zeroing out the components elsewhere. 

To implement this, we obtain $\hat{\mathbf{h}}_{k, \text{coarse}}$ from $\hat{\mathbf{y}}_{k,\text{ctr}}$, compute the 2D power spectrum of the reshaped matrix $\mathbf{H}^\text{unvec}_{k, \text{coarse}}$, smooth it with a small average window, and subtract a noise threshold (the mean intensity plus three standard deviations of the averaged spectrum). Kadane's algorithm \cite{Bentley84} is then applied along each dimension to find the energy-concentrated region, which identifies a rectangular spectral bounding box. Setting the frequency components outside this box to zero yields the coarse spatial response estimate $\hat{\mathbf{h}}_{k,\text{ctr}}$.

\subsection{Spatial Parameter Estimation Based on Phase Difference}
To estimate the spatial parameters $\bar{\mathbf{k}}_k$ and $\bar{\mathbf{Q}}_k$, a phase-difference operation is subsequently applied to the coarse spatial response $\hat{\mathbf{h}}_{k,\text{ctr}}$. First, we extract the reference sub-matrix $\mathbf{\hat{H}}_{k,\text{ctr}}^{(\text{unvec},1)} = [\mathbf{\hat{H}}_{k,\text{ctr}}^{\text{unvec}}]_{1{:}N_y{-}\Delta_y, \, 1{:}N_z{-}\Delta_z}$ and its spatially shifted counterpart $\mathbf{\hat{H}}_{k,\text{ctr}}^{(\text{unvec},2)} = [\mathbf{\hat{H}}_{k,\text{ctr}}^{\text{unvec}}]_{1{+}\Delta_y{:}N_y, \, 1{+}\Delta_z{:}N_z}$, whose spatial indices are offset from the reference by a displacement of $(\Delta_y, \Delta_z)$. To eliminate the first-order linear phase terms and extract the quadratic coefficients, the element-wise conjugate multiplication of the two sub-matrices is performed:
\begin{equation}
\begin{aligned}
    [\bm{\Psi}_1]_{n_y, n_z} &= \left([\mathbf{\hat{H}}_{k,\text{ctr}}^{(\text{unvec},1)}]_{n_y, n_z}\right)^* [\mathbf{\hat{H}}_{k,\text{ctr}}^{(\text{unvec},2)}]_{n_y, n_z} \\
    &= |a_k|^2 e^{j 2\pi (f_{1,y} \delta_{n_y} + f_{1,z} \delta_{n_z} + \Phi_1)} + [\mathbf{N}_1]_{n_y, n_z}
\end{aligned}
\end{equation}
where $f_{1,y} = -[\bar{\mathbf{Q}}_k]_{1,1} \Delta_y - [\bar{\mathbf{Q}}_k]_{1,2} \Delta_z$ and $f_{1,z} = -[\bar{\mathbf{Q}}_k]_{2,2} \Delta_z - [\bar{\mathbf{Q}}_k]_{1,2} \Delta_y$ are the differential frequencies. Here, $\Phi_1$ is a constant phase term independent of the spatial indices, and $\mathbf{N}_1[n_y, n_z]$ denotes the noise term. 

Similarly, we extract another two cross-shifted sub-matrices $\mathbf{\hat{H}}_{k,\text{ctr}}^{(\text{unvec},3)}= [\mathbf{\hat{H}}_{k,\text{ctr}}^{\text{unvec}}]_{1{:}N_y{-}\Delta_y,\,1{+}\Delta_z{:}N_z}$ and  $\mathbf{\hat{H}}_{k,\text{ctr}}^{(\text{unvec},4)} = [\mathbf{\hat{H}}_{k,\text{ctr}}^{\text{unvec}}]_{1{+}\Delta_y{}:N_y, 1{}:N_z{-}\Delta_z} $. Their element-wise conjugate multiplication yields:
\begin{equation}
\begin{aligned}
    [\bm{\Psi}_2]_{n_y, n_z} &= \left([\mathbf{\hat{H}}_{k,\text{ctr}}^{(\text{unvec},3)}]_{n_y, n_z}\right)^* [\mathbf{\hat{H}}_{k,\text{ctr}}^{(\text{unvec},4)}]_{n_y, n_z} \\
    &= |a_k|^2 e^{j 2\pi (f_{2,y} \delta_{n_y} + f_{2,z} \delta_{n_z} + \Phi_2)} + [\mathbf{N}_2]_{n_y, n_z}
\end{aligned}
\end{equation}
where the corresponding differential frequencies are $f_{2,y} = -[\bar{\mathbf{Q}}_k]_{1,1} \Delta_y + [\bar{\mathbf{Q}}_k]_{1,2} \Delta_z$ and $f_{2,z} = [\bar{\mathbf{Q}}_k]_{2,2} \Delta_z - [\bar{\mathbf{Q}}_k]_{1,2} \Delta_y$. By performing two-dimensional spectral peak searching on $\mathbf{R}_1$ and $\mathbf{R}_2$ to estimate these four differential frequencies, the curvature parameters can be estimated by equations below:
\begin{equation}\label{eq:quadratic params}
\begin{aligned}
    [\hat{\bar{\mathbf{Q}}}_k]_{1,1} &= \frac{-f_{1,y} - f_{2,y}}{2\Delta_y} \\
    [\hat{\bar{\mathbf{Q}}}_k]_{2,2} &= \frac{-f_{1,z} + f_{2,z}}{2\Delta_z} \\
    [\hat{\bar{\mathbf{Q}}}_k]_{1,2} &= [\hat{\bar{\mathbf{Q}}}_k]_{2,1} = \frac{1}{2} \left( \frac{-f_{1,y} + f_{2,y}}{2\Delta_z} + \frac{-f_{1,z} - f_{2,z}}{2\Delta_y} \right)
\end{aligned}
\end{equation}

The spatial lags $\Delta_y$ and $\Delta_z$ are carefully selected to ensure the estimation accuracy of $\bar{\mathbf{Q}}_k$. On one hand, increasing the spatial lags reduces the size of sub-matrices $\mathbf{\hat{H}}_{k,\text{ctr}}^{(\text{unvec},1)}$ through $\mathbf{\hat{H}}_{k,\text{ctr}}^{(\text{unvec},4)}$, which degrades the coherent accumulation gain and increases the estimation errors of the differential frequencies. On the other hand, according to (\ref{eq:quadratic params}), the ratio of the estimation errors of $\bar{\mathbf{Q}}_k$ to those of the differential frequencies is positively correlated with the inverse of the spatial lags. Thus, larger spatial lags makes the parameter estimation more robust against frequency errors. To reach an optimal balance between these two competing effects, we set $\Delta_y = \lfloor N_y/3 \rceil$ and $\Delta_z =\lfloor N_z/3 \rceil$ in our algorithm, where the operatior $\lfloor \cdot \rceil$ denotes rounding to the nearest integer.

Due to the noise in the phase difference matrices $\bm{\Psi}_1$ and $\bm{\Psi}_2$, false peaks are prone to appear in their 2D pseudo-spectra. To mitigate the interference from these false peaks and accurately extract the genuine differential frequencies for solving $\bar{\mathbf{Q}}_k$, we propose a robust peak-searching algorithm based on physical feasible regions and dimensionality reduction:
\begin{itemize}
    \item \textit{Delineating the Physical Feasible Region}: When evaluating the spectral intensities of $\bm{\Psi}_1$ and $\bm{\Psi}_2$, we first exclude frequency combinations that violate physical laws. Based on our assumptions regarding the convexity of the reflecting surfaces and their distances to the array center, it can be deduced from Section \ref{sec:AWC} that the curvature matrix must satisfy the following constraints:
    \begin{equation}
        \bar{\mathbf{Q}}_k  \succeq 0, \quad \text{and} \quad \lambda_{\max}(\bar{\mathbf{Q}}_k) \le \frac{d_{\text{ant}}^2}{\lambda r_{\min}}
    \end{equation}
    Furthermore, by substituting this feasible region of physical parameters into the differential frequency mapping relations (\ref{eq:quadratic params}), the upper and lower bounds for each frequency coordinate $f_{i,y}$ and $f_{i,z}$ for $i=1,2$ can be calculated. This restricts the peak-searching space to a bounded rectangular frequency region.
    \item \textit{Dimensionality Reduction via Constraints}:  According to their expressions, the differential frequencies strictly satisfy the following linear constraint:
    \begin{equation}
        \Delta_y f_{1,y} - \Delta_z f_{1,z} = \Delta_y f_{2,y} + \Delta_z f_{2,z} = V
    \end{equation}
    This constraint implies that, given a specified intercept $V$, the corresponding differential frequency points in the power spectra of $\bm{\Psi}_1$ and $\bm{\Psi}_2$ are confined to two lines with opposite slopes. The true intercept $V^*$ corresponds to the specific lines that simultaneously intersect the genuine spectral peaks in both spectra. To guarantee this joint peak intersection, we traverse $V$ through its range and maximize the minimum of the local peak intensities along its corresponding lines, thereby identifying the true intercept $V^*$ and its associated differential frequencies. The range of the intercept $V$ can be determined by the aforementioned boundaries of the differential frequencies, while its discrete step size is determined by the frequency sampling density of the spectra.
\end{itemize}

By compensating for the quadratic phase of $\mathbf{h}_{k,\text{ctr}}$ with the estimated curvature, the direction parameters $\bar{\mathbf{k}}_k$ are extracted from the residual single-frequency signal. We define a phase compensation matrix $\mathbf{S}_k$, whose elements are given by:
\begin{equation}
    [\mathbf{S}_k]_{n_y, n_z} = e^{-j\pi\left(\mathbf{n}_{y,z}^T \hat{\bar{\mathbf{Q}}}_k \mathbf{n}_{y,z} \right)}
\end{equation}
It is then diagonalized as $\mathbf{D}_{\mathbf{S}_k} = \text{diag}(\text{vec}(\mathbf{S}_k))$. With the diagonalized matrix, the original channel vector can then be factored as $\mathbf{h}_{k,\text{ctr}} = \mathbf{D}_{\mathbf{S}_k} \hat{\mathbf{h}}_{k,\text{ctr}}^{\text{single}}$. Here, $\hat{\mathbf{h}}_{k,\text{ctr}}^{\text{single}}$ approximates a single-frequency vector that shares the identical direction parameters with $\mathbf{h}_{k,\text{ctr}}$. Its quadratic phase terms corresponds to the curvature estimation errors.

To recover the single-frequency vector accurately, we revisit the observation model provided by (\ref{eq:center freq observation}). By substituting the decomposition of $\mathbf{h}_k$, it is rewritten as:
\begin{equation}
\hat{\mathbf{y}}_{k,\text{ctr}} = \mathbf{W}\mathbf{D}_{\mathbf{S}_k} \hat{\mathbf{h}}_{k,\text{ctr}}^{\text{single}} + \mathbf{n}_k
\end{equation}
Note that the equivalent measurement matrix, defined as $\tilde{\mathbf{W}}_k = \mathbf{W}\mathbf{D}_{\mathbf{S}_k}$, is also an SRFT matrix. By substituting $\mathbf{W}$ with $\tilde{\mathbf{W}}_k$, we can recover $\hat{\mathbf{h}}_{k,\text{ctr}}^{\text{single}}$ following the procedure described in the previous subsection. Finally, the estimate $\hat{\bar{\mathbf{k}}}_k$ can be obtained by applying 2D FFT to the unvectorized $\hat{\mathbf{h}}_{k,\text{ctr}}^{\text{single}}$ and searching for the global peak. In practice, we directly apply 2D FFT to $\mathbf{S}_k^* \odot \mathbf{\hat{H}}_{k,\text{ctr}}^{\text{unvec}}$ and identifying the peak location, which is equivalent to the above procedure.

\subsection{Parameter Refinement and Channel Reconstruction}

The spatial parameter set $(\hat{\bar{\mathbf{k}}}_k, \hat{\bar{\mathbf{Q}}}_k)$ estimated above has already converged to the neighborhood of the true values. Based on the Levenberg-Marquardt (LM) algorithm \cite{Marquardt63}, we perform a two-stage fine-tuning on these parameters and simultaneously obtain the path coefficient $a_k$.

In the first stage, we neglect the beam split effect and assume that the multipath components have been perfectly decoupled by the RELAX algorithm, aiming to coarsely tune the spatial parameters with low computational complexity. Under these assumptions, fitting the spatial parameters over the wideband observation $\mathbf{Y}$ is equivalent to fitting them over the RELAX spatial observation $\hat{\mathbf{y}}_{k,\text{ctr}}$ for each path. This simplifies the objective function to the following form:
\begin{equation}
\min_{\hat{a}_k, \hat{\bar{\mathbf{k}}}_k, \hat{\bar{\mathbf{Q}}}_k} \Vert \hat{\mathbf{y}}_{k,\text{ctr}} - \hat{a}_k \mathbf{Wc}_k(\hat{\bar{\mathbf{k}}}_k, \hat{\bar{\mathbf{Q}}}_k) \Vert^2_2
\end{equation}

In the second stage, we take the beam split effect into account and jointly refine the distance parameters provided by RELAX alongside the spatial parameters from the first stage. Initialized with the parameters obtained thus far, we optimize the following objective function:
\begin{equation}
\min_{\bm{\Theta}} \left\Vert \mathbf{Y} - \mathbf{W} \sum_{k=1}^{K_\text{path}}\hat{a}_k \tilde{\mathbf{H}}_k(\hat{\bar{S}}_k, \hat{\bar{\mathbf{k}}}_k, \hat{\bar{\mathbf{Q}}}_k) \right\Vert^2_F
\end{equation}
where $\bm{\Theta} = \{\hat{a}_k, \hat{\bar{S}}_k, \hat{\bar{\mathbf{k}}}_k, \hat{\bar{\mathbf{Q}}}_k\}_{k=1}^{K_\text{path}}$ denote the set of parameters for all paths, and $\tilde{\mathbf{H}}_k$ denotes the reconstructed wideband sub-channel for the $k$-th path, which is obtained by substituting the corresponding parameters into (\ref{eq:general path expression}) and aggregating the responses over all subcarriers. Given the high-quality initial values yielded by RELAX and the first-stage optimization, the second stage requires only a few iterations to converge. 

To reduce the computational complexity of the second-stage refinement, we adopt a Taylor-approximated Jacobian matrix within the LM algorithm. In each LM iteration, the complex path gains $\{\hat{a}_k\}_{k=1}^{K_\text{path}}$ are first obtained via LS between the observation $\mathbf{Y}$ and each reconstructed sub-channel $\mathbf{W}\tilde{\mathbf{H}}_k$. With these gains held constant, the Jacobian of the residual matrix $\mathbf{R} = \mathbf{Y} - \mathbf{W} \sum_{k=1}^{K_\text{path}}\hat{a}_k \tilde{\mathbf{H}}_k$ with respect to a spatial parameter $p_k$ is provided by:
\begin{equation}
\frac{\partial \mathbf{R}}{\partial p_k} = j 2\pi\hat{a}_k \mathbf{W} \left(  \left(\mathbf{u}_{p_k}\bar{\mathbf{f}}^T\right) \odot \tilde{\mathbf{H}}_k \right)
\end{equation}
where $p_k$ represents $\hat{\bar{S}}_k$, or any scalar element within $\hat{\bar{\mathbf{k}}}_k$ and $\hat{\bar{\mathbf{Q}}}_k$. $[\bar{\mathbf{f}}]_m = 1+\frac{\Delta f}{f_c}\delta_m$, and $\mathbf{u}_{p_k}$ is a constant spatial coordinate vector associated with parameter $p_k$. By regarding the $m$-th column of $\mathbf{P}_k = \left(\mathbf{u}_{p_k}\bar{\mathbf{f}}^T\right) \odot \tilde{\mathbf{H}}_k$ as a vector dependent on $\delta_m$, it can be linearly approximated via a Taylor expansion as $\mathbf{p}_k + \delta_m \Delta\mathbf{p}_k$. Therefore, for the $m$-th column of the Jacobian, we have the approximation:
\begin{equation}
\left[\frac{\partial \mathbf{R}}{\partial p_k}\right]_{1:N,m} \approx j 2\pi\hat{a}_k \left( \mathbf{Wp}_k+\delta_m  \mathbf{W}\Delta\mathbf{p}_k\right)
\end{equation}
Since $\mathbf{Wp}_k$ and $\Delta \mathbf{Wp}_k$ are independent of the subcarrier index and only need to be evaluated once per iteration, this procedure is significantly faster than calculating the exact Jacobian directly, while maintaining high approximation accuracy under moderate beam split conditions.

Finally, the overall wideband channel estimate $\hat{\mathbf{H}}$ is obtained by superimposing all path components evaluated across all subcarriers via (\ref{eq:general path expression}).

\subsection{Complexity Analysis}
The computational complexity of each stage in our proposed algorithm is summarized as follows. 

The following procedures are executed per path, accumulating to $K_\text{path}$ iterations in total:
\begin{itemize}
\item \textit{Recovery of} $\hat{\mathbf{h}}_{k,\text{ctr}}$: For each path, the left multiplication by $\mathbf{W}^H$ takes $\mathcal{O}(N \log N)$. Locating the most energy-concentrated rectangular region via the 1D Kadane's algorithm requires $\mathcal{O}(N_y + N_z)$. The projection operation (zeroing out components outside the region) takes $\mathcal{O}(N)$.
\item \textit{Parameter Estimation}: The element-wise multiplication has a complexity of $\mathcal{O}(N)$, and computing the intensity spectra of $\bm{\Psi}_1$ and $\bm{\Psi}_2$ takes $\mathcal{O}(N \log N)$. As the intercept $V$ is discretely traversed, the two corresponding lines sweep across the entire 2D spectral grids without omission or overlap. Therefore, optimizing $V$ is equivalent to performing a single linear scan over the two spectra, yielding a complexity of $\mathcal{O}(N)$. Recovering $\hat{\mathbf{h}}_{k,\text{ctr}}^{\text{single}}$ and searching for peaks in its spectrum requires $\mathcal{O}(N \log N)$. 
\item \textit{First-stage Parameter Refinement}:
For LM algorithm, the most time-consuming step per iteration is to calculate the Jacobian $\mathbf{J}_k$ which can be efficiently performed by SRFT with a complexity of $\mathcal{O}(N \log N)$. 
\end{itemize}
The computational complexity of the procedures executed once in the proposed algorithm is summarized as follows:
\begin{itemize}
    \item \textit{RELAX}: Given the observation data scale of $P N_{\text{RF}}$ spatial samples across $M$ subcarriers, the RELAX algorithm extracts distance parameters and recovers sub-channels via iterative interference cancellation and spectral search. Its complexity is $\mathcal{O}(K_\text{path}^2 P N_{\text{RF}} M \log M)$ \cite{Li95}.
    \item \textit{Second-stage Parameter Refinement}: During the joint optimization, evaluating the Taylor-approximated spatial basis vectors $\mathbf{Wp}_k$ and $\mathbf{W}\Delta\mathbf{p}_k$ takes $\mathcal{O}(N \log N)$ operations per path. Subsequently, expanding this approximation across all $M$ subcarriers requires $\mathcal{O}(P N_{\text{RF}} M)$ operations. Thus, constructing the complete Jacobian matrix incurs a complexity of $\mathcal{O}(K_\text{path} N \log N + K_\text{path} P N_{\text{RF}} M)$. Additionally, the complexity of the estimation for complex path gains through LS is $\mathcal{O}(K_\text{path}^2 P N_{\text{RF}} M)$.
    \item \textit{Channel Reconstruction}: The final wideband physical channel is reconstructed by analytically aggregating the responses of all $K_\text{path}$ paths across $N$ antennas and $M$ subcarriers. This yields a straightforward complexity of $\mathcal{O}(K_\text{path} N M)$.
\end{itemize}
In summary, considering that $P N_{\text{RF}} \ll N$ in hybrid precoding systems, the total complexity is dominated by the channel reconstruction, which is $\mathcal{O}(K_{\text{path}} N M)$. Any parameter-recovery-based channel estimation algorithm must reconstruct the channel path by path, meaning that our algorithm achieves the theoretical minimum complexity. 

Moreover, if only the channel parameters are required for downstream tasks, the channel reconstruction step can be omitted, thus the algorithm complexity becomes $\mathcal{O}(K_{\text{path}}N \log N + K_{\text{path}}^2P N_{\text{RF}} M \log M)$. This overall complexity is substantially lower than the typical $\mathcal{O}(K_\text{path} P N_{\text{RF}} M Q)$ complexity required by traditional matching-based algorithms, where the codebook size $Q$ is generally comparable to the massive number of antennas $N$. Consequently, the proposed algorithm guarantees a tremendous speed advantage over conventional approaches.

\section{Simulation Results} \label{sec:Simulation}
In this section, we begin by illustrating how wavefront anisotropy induces non-sparsity in the polar-domain channel representation. Subsequently, we validate the accuracy of the theoretical lower bound derived in Section \ref{sec:AWC} for the AWC-SWC similarity, and further investigate the impact of physical characteristics on the degree of wavefront anisotropy. Finally, we evaluate the effectiveness of the proposed algorithm. Since the topics discussed in Subsections A-C do not involve the frequency-domain characteristics of the channel, the simulations therein are conducted under narrowband configurations.
\subsection{Non-Sparsity of AWC in the Polar Domain}

\begin{figure}[htbp]
    \centering
    \includegraphics[width=\linewidth]{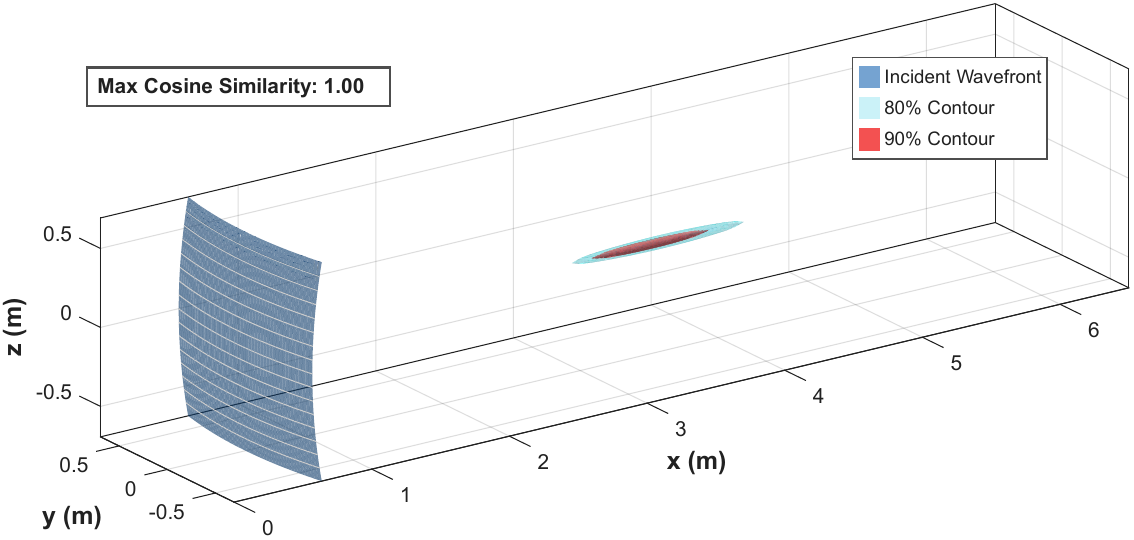} \\ 
    \includegraphics[width=\linewidth]{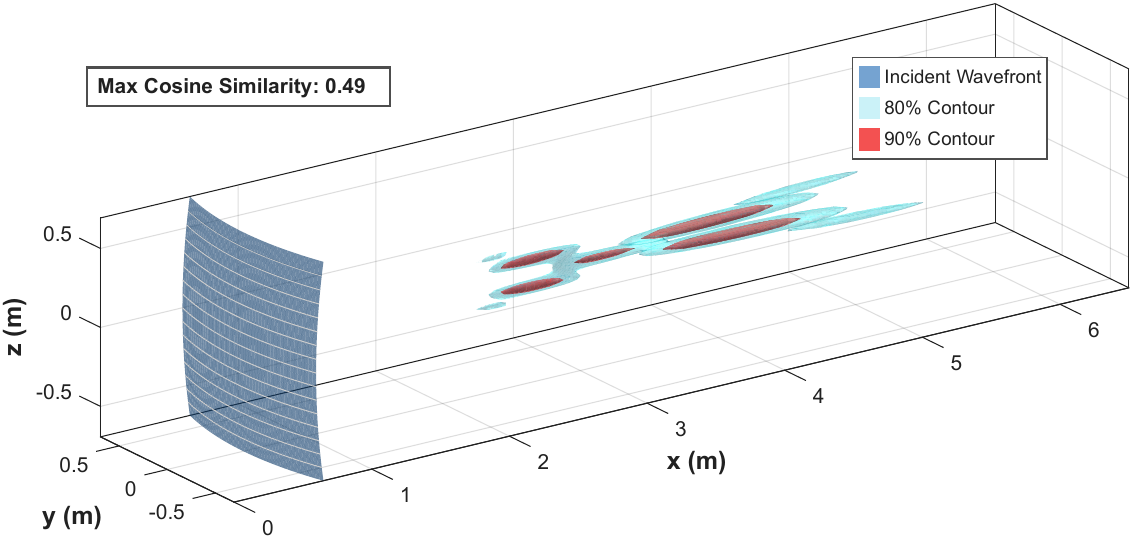}
    \caption{Spatial similarity distribution of (top) SWC and (bottom) AWC. For clarity, the BS array is omitted in the figure.}
    \label{fig:wavefront_comparison}
\end{figure}

In Fig. \ref{fig:wavefront_comparison}, we evaluate how accurately the channels corresponding to two distinct incident electromagnetic waves can be fitted by polar-domain steering vectors. Both waves incident normally on the array but exhibit different wavefront curvatures. For the AWC, the principal radii of curvature are 2.5 m and 4.5 m, with the principal directions aligned with the array axes. For the SWC, the radius of curvature is 3.5 m. The antenna array size is set to $64 \times 64$, the carrier frequency is 7.5 GHz, and the antenna spacing is a half-wavelength.

Fig. \ref{fig:wavefront_comparison} illustrates the polar-domain non-sparsity of the AWC by highlighting the maximum cosine similarity for both channels, along with the volumetric contours capturing 80\% and 90\% of this peak similarity. The spatial similarity distribution for the SWC is highly concentrated, reaching a maximum of 1.0 at $x = 3.5$ m along the propagation axis. This indicates that the SWC can be accurately represented by a single point source. In contrast, a single point source achieves a maximum cosine similarity of only about 0.5 with the AWC. Instead of concentrating at a single focal point, the spatial energy of the AWC diffuses into two distinct, orthogonal caustic line segments located at $x = 2.5$ m and $x = 4.5$ m, which correspond to its two principal radii of curvature.

\subsection{Validation of the Similarity Lower Bound}
\begin{figure}[!t]
\centering
\includegraphics[width=\linewidth]{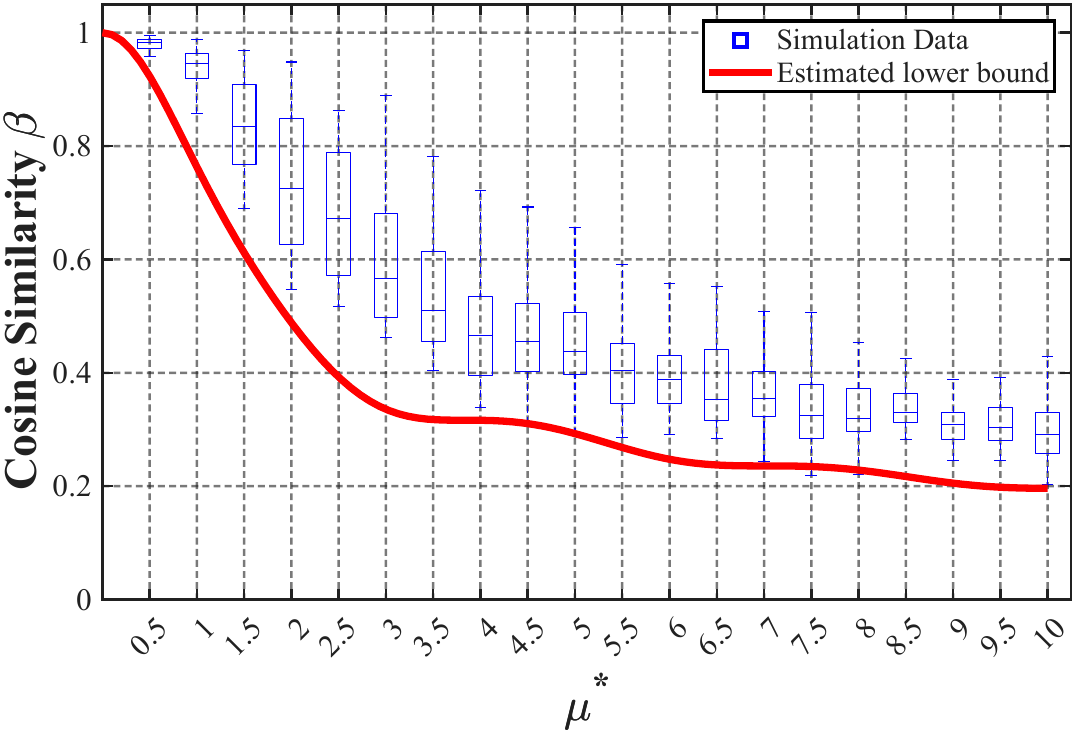}
\caption{Comparison between the theoretical lower bound of the maximum cosine similarity and the statistical distribution of 1000 randomized numerical search results.}
\label{img:AWC_vs_SWC_Cosine_Similarity}
\end{figure}

To verify the correctness of (\ref{eq:cossim expression}), we randomly generated 1000 AWC steering vectors with different wavefront shapes and array scales, and then determined the best-matching spherical wave steering vector along with the optimal cosine similarity via exhaustive search in continuous parameter domains. To ensure statistical reliability, we controlled the principal curvature differences such that the test samples were uniformly distributed across the region of interest $\mu^* \in [0, 10]$. This region was then divided into 20 sub-intervals (i.e., 50 samples per sub-interval). The statistical distribution of the cosine similarity within each sub-interval is compared with the theoretical lower bound given by (\ref{eq:cossim expression}), as plotted in Fig. \ref{img:AWC_vs_SWC_Cosine_Similarity}. It can be observed that the statistical boxplots closely follow the estimated theoretical lower bound, which validates the accuracy of our derived expression. 

\subsection{Impact of Propagation Characteristics on AWC Anisotropy}

In this subsection, we design the following experiments to investigate the impact of various propagation characteristics:
\begin{itemize}
    \item Varying the distance from the scatterer to the BS array center, while keeping the BS-scatterer relative direction and the UE-scatterer relative position fixed.
    \item Varying the distance from the UE to the scatterer, while keeping the UE-scatterer relative direction and the BS-scatterer relative position fixed.
    \item Varying the antenna array scale and the carrier wavelength simultaneously, while keeping the physical array aperture constant.
    \item Varying the antenna array scale to increase its diagonal length, while keeping the carrier wavelength constant.
    \item Varying the surface curvature of the scatterer.
\end{itemize}
Unless otherwise specified in Table \ref{tab:nmse_results}, the default parameters for the experiments are set as follows: the antenna array size is $128 \times 128$, and the carrier frequency is $15$ GHz. The center of the BS array is positioned at $[0, 0, 10]^T$ m, while the UE is located at $[30, 0, 1.5]^T$ m. The array has a downtilt angle of $\phi = \pi/18$. The propagation scenario features a single cylindrical scatterer with principal curvatures of $4\text{ m}^{-1}$ and $0\text{ m}^{-1}$, which are aligned with the $x$- and $z$-axes, respectively. For each generated AWC, we identify the optimal SWC that minimizes the normalized mean square error (NMSE) between their respective channel vectors. This minimized NMSE is subsequently employed as the quantitative metric to characterize the degree of wavefront anisotropy in the AWC.

\begin{table}[t]
  \centering
  \caption{NMSE of SWC-Fitting Across Various Physical Configurations}
  \label{tab:nmse_results}
  \resizebox{\linewidth}{!}{
  \begin{tabular}{|c|ccccc|}
    \hline
    \multicolumn{6}{|c|}{\textbf{Scatterer-BS Distance}} \\
    \hline
    $r$ (m) & 5 & 8 & 10 & 15 & 20 \\
    NMSE & 8.94e-01 & 8.06e-01 & 6.99e-01 & 3.48e-01 & 1.78e-01 \\
    \hline
    \multicolumn{6}{|c|}{\textbf{UE-Scatterer Distance}} \\
    \hline
    $d$ (m) & 2.0 & 4.0 & 6.0 & 8.0 & 22.8 \\
    NMSE & 4.33e-02 & 1.42e-01 & 2.43e-01 & 3.30e-01 & 6.30e-01 \\
    \hline
    \multicolumn{6}{|c|}{\textbf{Array Scale \& Frequency}} \\
    \hline
    $N$, $f_c$ & $32^2$, 3.8G & $64^2$, 7.5G & $128^2$, 15.0G & $192^2$, 22.5G & $256^2$, 30.0G \\
    NMSE & 6.22e-02 & 2.27e-01 & 6.30e-01 & 8.29e-01 & 8.88e-01 \\
    \hline
    \multicolumn{6}{|c|}{\textbf{Array Diagonal Length}} \\
    \hline
    $D$ (m) & 0.23 & 0.45 & 0.91 & 1.81 & 3.62 \\
    NMSE & 2.49e-04 & 4.01e-03 & 6.23e-02 & 6.30e-01 & 9.85e-01 \\
    \hline
    \multicolumn{6}{|c|}{\textbf{Scatterer Curvature}} \\
    \hline
    $R_\text{s}^{-1}$ (m$^{-1}$) & 0.0 & 0.5 & 2.0 & 4.0 & 10.0 \\
    NMSE & 1.06e-12 & 4.17e-01 & 5.95e-01 & 6.30e-01 & 6.50e-01 \\
    \hline
  \end{tabular}
  }
\end{table}

The simulation results in Table \ref{tab:nmse_results} reveal how various physical characteristics influence the degree of wavefront anisotropy:

\begin{itemize}
    \item \textit{Scatterer-to-BS distance:} A larger distance mitigates wavefront anisotropy. As the scattered wave propagates, its wavefront naturally flattens, reducing the disparity between its two principal curvatures and thereby making it converge toward the SWC model.
    \item \textit{UE-to-scatterer distance:} A shorter distance similarly diminishes anisotropy. A UE positioned closer to the scatterer induces a larger incident curvature, imparting a greater initial curvature to the reflected wave. For an anisotropic wavefront, the principal direction with the larger curvature flattens more rapidly during propagation. Consequently, a larger initial curvature accelerates this differential flattening process, which reduces the wavefront anisotropy upon arrival at the BS.
    \item \textit{Carrier frequency and array scale:} For a fixed physical aperture, increasing both the carrier frequency and the array scale renders the channel phase more sensitive to the precise wavefront geometry, thereby amplifying the observable effects of wavefront anisotropy.
    \item \textit{Array diagonal length:} A larger diagonal length exacerbates the anisotropic effect. The curvature mismatch caused by approximating an anisotropic wavefront with an SWC introduces more severe phase deviations at the antennas located further from the array center.
    \item \textit{Scatterer surface curvature:} A larger surface curvature inherently increases the initial anisotropy of the scattered wave, widening its deviation from the SWC model. Notably, when the scatterer's principal curvatures are both zero (i.e., a reflection plane), the reflection becomes purely specular. This scenario is geometrically equivalent to an electromagnetic wave emitted from a virtual UE source directly reaching the BS, rendering the AWC identical to an SWC.
\end{itemize}

\subsection{Performance of the Proposed Algorithm}
In this section, we evaluate the NMSE curves of the proposed algorithm against several baselines and theoretical bounds under wideband channel conditions. The system bandwidth is set to $B = 100$ MHz. Both the number of RF chains $N_{\text{RF}}$ and the number of pilot symbols $P$ are set to $16$, and the number of paths is $K_{\text{path}} = 6$. The minimum distance from a scatterer to the BS is calculated as $r_{\min} = 1.7$ m. The remaining parameters inherit the default settings from the previous subsection. To compare the algorithmic performance under varying conditions, 100 different propagation scenarios and their corresponding channels are generated for both the AWC and SWC cases, with the average NMSE serving as the performance metric.

The channels to be estimated are generated according to the following procedure. For AWC generation, we consider exclusively single-bounce reflections, intentionally omitting the LOS path, ground reflections, and multiple-bounce reflections. To generate the $K_{\text{path}}$ paths, $K_{\text{path}}$ distinct scatterers are randomly generated between the BS array and the UE, with the distance from each scatterer to the BS being no less than $r_\text{min}$. Their principal curvatures are identical to those defined in the previous subsection; however, their principal directions are randomly chosen, and their surface normals are dictated by the law of reflection. Specifically, the total propagation distance via the $k$-th scatterer is set recursively as $d_k = d_{k-1} + \Delta d_k$, where $d_0$ represents the distance between the BS and the UE. The distance increment $\Delta d_k$ is uniformly drawn from the interval $[0.8 c_0 / B, 1.2 c_0 / B]$. Furthermore, the phase and amplitude of the reflection coefficient at each scatterer surface are uniformly distributed over $[0, 2\pi)$ and $[0.2, 0.8]$, respectively. Finally, the path loss for each multipath component is computed based on the formulations in \cite{Deschamps72}, and the propagation directions are governed by the law of reflection. For SWC generation, the procedure is identical to that of the AWC, except that both curvatures of the  scatterers are set to zero. 

The algorithms for comparison are listed as follows:
\begin{itemize}
\item \textit{MACAW}: Our proposed algorithm.
\item \textit{SWC-Fitting}: This baseline assumes that the sub-channels corresponding to all individual paths are perfectly known. For each sub-channel, it performs an exhaustive search over the direction and distance parameters to find the SWC basis that minimizes MSE, followed by a LS estimation using the matched channel bases. The results establish the performance upper bound for 
channel estimation algorithms based on SWC matching and assuming the exact number of paths is known a priori.

\item \textit{CRLB}: The Cramér-Rao lower bound under the specific noise level. CRLB(AWC) and CRLB(SWC) denote the theoretical bounds computed based on the AWC and SWC multipath channel models, respectively.
\item \textit{P-SIGW-UPA}: Assuming the traditional SWC model, this approach first utilizes frequency domain information to determine the distance of each path and its associated wavefront curvature. Subsequently, it generates a near-field codebook based on these parameters and estimates the channel via matching-based sparse recovery \cite{Chen25}.
\end{itemize}

\begin{figure}[htbp]
    \centering
    \includegraphics[width=\linewidth]{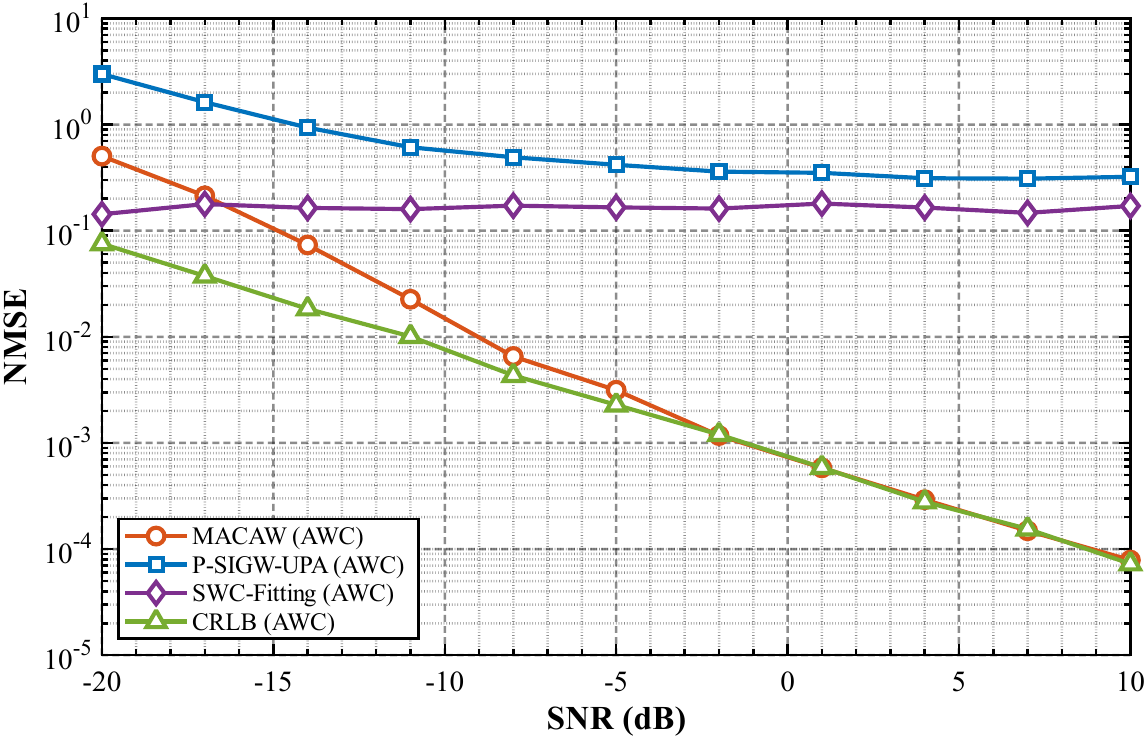} \\ 
    \includegraphics[width=\linewidth]{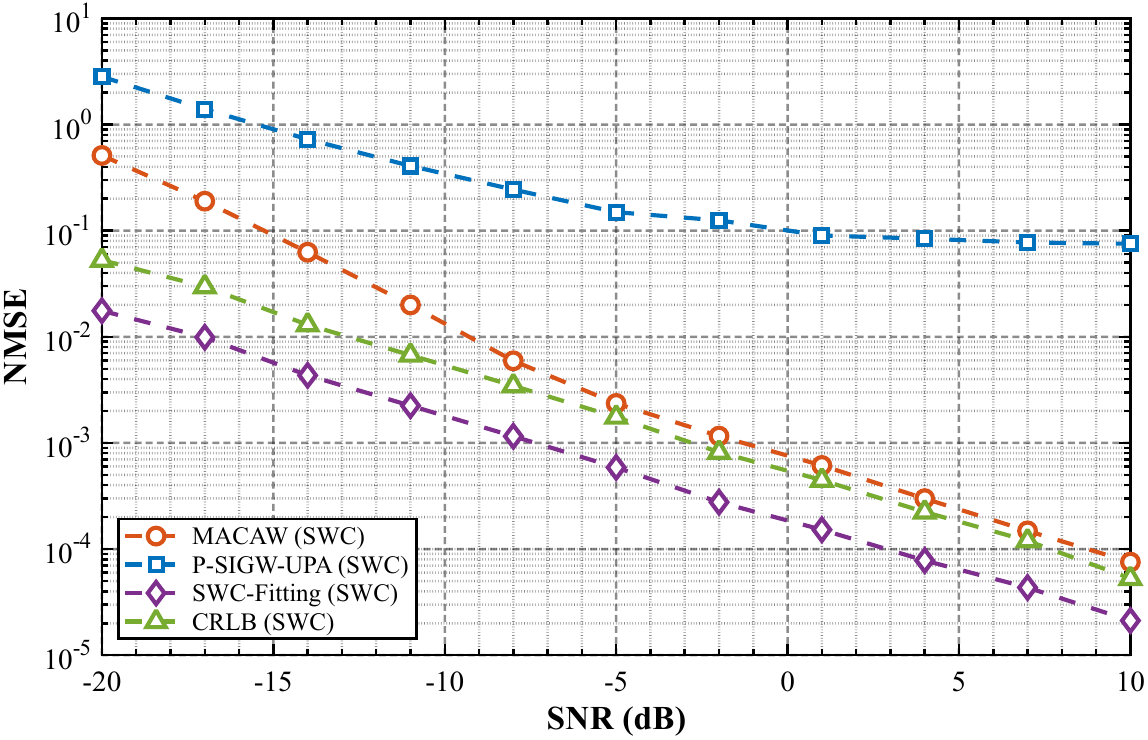}
    \caption{NMSE curves of compared algorithms in multipath (top) AWC and (bottom) SWC.}
    \label{fig:NMSE_vs_SNR}
\end{figure}
From Fig. \ref{fig:NMSE_vs_SNR}, the proposed algorithm approaches CRLB at high SNRs. At low SNRs, performance experiences degradation caused by the noise amplification of the phase-difference method but maintains robustness, with 99\% accuracy at -10 dB. SWC-Fitting results indicate that the actual channel is not accurately modeled by SWC due to its anisotropic wavefront, resulting in a model error over 10\%. This model mismatch limits the high-SNR precision of the SWC-based P-SIGW-UPA algorithm. Additionally, the gap between P-SIGW-UPA and SWC-Fitting suggests that SWC-based strategies may fail to find the best-fitting spherical wave for AWC, further degrading its estimation accuracy.

Simultaneously, in the SWC model, the curvature matrix is determined solely by a single distance parameter once the arrival directions are given. Since the proposed algorithm is developed based on the AWC model, it inherently introduces two redundant parameters. This over-parameterization leads to a slight overfitting effect, rendering its estimation accuracy at high SNRs marginally inferior to the CRLB. Furthermore, a cross-comparison between AWC and SWC reveals that the proposed algorithm exhibits identical performance across both AWC and SWC scenarios. Meanwhile, the SWC-Fitting perfectly matches the channel under the SWC assumption; because it utilizes the exact prior knowledge of individual paths, its estimation error falls below the CRLB. With the modeling mismatch eliminated, The P-SIGW-UPA algorithm yields an improved performance in the SWC, yet a substantial gap to the theoretical performance bound remains. This is because it determines the curvature by estimating the propagation distance of each path, and the accuracy of curvature estimation decreases when the number of observations is small or the propagation distance is large \cite{Chen25}.

\section{Conclusion} \label{sec:Conclusion}
This paper investigates wideband multipath channels exhibiting wavefront anisotropy in ELAA systems. We first establish a parameterized model for the AWC and derive a lower bound on its similarity to SWC. This lower bound not only provides a theoretical criterion for model selection but also reveals the physical factors affecting the degree of channel wavefront anisotropy. Through numerical simulations, we demonstrate the polar-domain non-sparsity of the AWC, validate the accuracy of the derived lower bound, and analyze the impact of diverse physical characteristics on wavefront anisotropy. Simulation results reveal that the energy of the AWC is not concentrated at a single focal point in the polar domain; instead, it disperses across two caustic lines determined by the principal curvatures and principal directions of the wavefront. Specifically, wavefront anisotropy is exacerbated by a shorter scatterer-to-BS distance, a larger UE-to-scatterer distance, a shorter carrier wavelength, a larger array diagonal length, and a greater scatterer surface curvature.

Furthermore, we propose MACAW, an efficient algorithm for estimating wideband multipath AWCs with a computational complexity significantly lower than that of existing methods. This approach first leverages frequency-domain information to separate multipath components with distinct delays and extract the propagation distance of each path. Subsequently, it utilizes phase differencing to estimate the direction and curvatures of each path, followed by a low-complexity optimizer to further refine the estimated parameters. Simulation results demonstrate that while existing channel estimation algorithms suffer from severe performance degradation under the AWC due to model mismatch, our proposed algorithm approaches the CRLB at SNR as low as -5 dB. Although the performance of the proposed algorithm experiences a decline at lower SNRs due to the noise amplification effect inherent in phase differencing, it still maintains a 99\% estimation accuracy at an SNR of -10 dB. Finally, when operating in SWC scenarios, the overparameterization of our algorithm introduces only a marginal performance degradation relative to the CRLB.

{\appendix[Derivation of the cosine similarity lower bound]\label{sec:Appendix}
This appendix first derives an integral approximation for the lower bound of the maximum cosine similarity between a specific AWC steering vector and a SWC steering vector. Subsequently, the value of the maximum is determined.
\subsection{Integral Approximation of the Cosine Similarity}
According to Remark 2 in Section \ref{sec:AWC}, an arbitrary SWC steering vector $\mathbf{b}$ is also an AWC steering vector, thus it can be represented as $\mathbf{c}(\bar{\mathbf{k}}_{\text{sph}},\bar{\mathbf{Q}}_{\text{sph}})$. Substituting this representation into the expression of $\beta$ yields:
\begin{equation}
    \beta = \max_{\mathbf{k}_{\text{sph}}, \bar{\mathbf{Q}}_{\text{sph}}} \left|\frac{1}{N} \sum_{i} [\mathbf{c}\big(\bar{\mathbf{k}} - \bar{\mathbf{k}}_{\text{sph}}, \bar{\mathbf{Q}} - \bar{\mathbf{Q}}_{\text{sph}}\big)]_i\right|.
\end{equation}
By setting $\bar{\mathbf{k}}_{\text{sph}} = \bar{\mathbf{k}}$, we can obtain a lower bound for $\beta$. Let $R_{\text{sph}}$ denote the curvature radius of the SWC steering vector $\mathbf{b}$. The corresponding effective curvature matrix can then be expressed as $\bar{\mathbf{Q}}_{\text{sph}} = \frac{d_{\text{ant}}^2}{\lambda R_{\text{sph}}}\mathbf{P}^T\mathbf{P}$. Consequently, the lower bound is given by
\begin{equation}
    \beta \geq \max_{R_{\text{sph}}} \left|\frac{1}{N} \sum_{i} \left[\mathbf{c}\left(\mathbf{0}, \frac{d_{\text{ant}}^2}{\lambda} \mathbf{P}^T\Big(\mathbf{Q}_{\text{BS}} - \frac{1}{R_{\text{sph}}}\mathbf{I}_2\Big)\mathbf{P}\right)\right]_i\right|.
\end{equation}
Let $\mathbf{D} = \text{diag}([\frac{N_y}{2}, \frac{N_z}{2}]^T)$ and define the weighted curvature mismatch matrix as $\mathbf{W}(R_{\text{sph}}) = \frac{\pi d_{\text{ant}}^2}{\lambda}\mathbf{D}^T \mathbf{P}^T \big(\mathbf{Q}_{\text{BS}} - \frac{1}{R_{\text{sph}}} \mathbf{I}_2\big) \mathbf{P} \mathbf{D}$. Using a continuous integral to approximate the discrete summation, the lower bound is estimated by:
\begin{equation}\label{eq:square integral}
    \beta \geq \frac{1}{4}\left|\int_{[-1,1]^2} e^{-j(\mathbf{x}^T \mathbf{W} \mathbf{x})} d\mathbf{x}\right|.
\end{equation}
To simplify this expression, we convert the integration domain from the square $[-1,1]^2$ to the unit disk:
\begin{equation}
    \beta \geq \frac{t}{4}\left|\int_{\Vert\mathbf{x}\Vert_2^2 \leq 1} e^{-j(\mathbf{x}^T \mathbf{\Lambda} \mathbf{x})} d\mathbf{x}\right|.
\end{equation}
Here, $t$ is a positive scaling factor for compensating the similarity loss incurred by omitting the integral on the four corners of the square. The explicit expression of $t$ is derived at the end of this appendix.

Transforming the integration variables into polar coordinates with $\mathbf{x} = [r\cos\theta, r\sin\theta]^T$, the quadratic phase term evaluates to $\mathbf{x}^T \mathbf{\Lambda} \mathbf{x} = r^2 (\lambda_1 \cos^2\theta + \lambda_2 \sin^2\theta)$. By defining $S = \frac{1}{2}(\lambda_1 + \lambda_2)$ and $\mu = \frac{1}{2}|\lambda_1 - \lambda_2|$, we can rewrite this phase term using the double-angle formula as $\mathbf{x}^T \mathbf{\Lambda} \mathbf{x} = r^2(S \pm \mu \cos 2\theta)$. The integral over the unit disk then becomes:
\begin{equation}
    \beta \geq \frac{t}{4}\left|\int_0^1 e^{-jSr^2} \Big(\int_0^{2\pi}  e^{-j\mu r^2 \cos 2\theta}  d\theta\Big) rdr\right| .
\end{equation}
The inner integral $\int_0^{2\pi}  e^{-j\mu r^2 \cos 2\theta} d\theta$ is identical to $2\pi J_0(\mu r^2)$. By applying the substitution $u = r^2$, the expression simplifies  to the following one-dimensional integral:
\begin{equation}\label{eq:cossim}
    \beta \geq \frac{\pi}{4}t\left| \int_0^1 e^{jSu} J_0(\mu u) du \right|.
\end{equation}

\subsection{Maximization of the Integral}
To maximize the integral, one possible approach is to let $|S| = \mu$, and here is the explanation. Note that the integral can be interpreted as the Fourier transform of a truncated zero-order Bessel function. For a sufficiently large $\mu u$, the Bessel function can be asymptotically approximated by \cite{Balanis12}:
\begin{equation}
    J_0(\mu u) \approx \sqrt{\frac{2}{\pi \mu u}} \cos\left(\mu u - \frac{\pi}{4}\right).
\end{equation}

Since the amplitude envelope $\sqrt{2 / (\pi \mu u)}$ is a slowly varying function over the integration domain, the peak of the integral is determined by the highly oscillatory cosine term, indicating that $|S| = \mu$ to maximize $\beta$. Substituting this optimal condition back into the integral, the maximum peak value can be analytically evaluated as a closed-form expression \cite{Gradshteyn07}:
\begin{equation}
    \beta \geq \frac{\pi}{4}t\sqrt{J_0^2(\mu_m) + J_1^2(\mu_m)}, \quad m \in \{1, 2\}.
\end{equation}
where $\mu_1$ and $\mu_2$ correspond to the absolute difference of the eigenvalues of $\mathbf{W}$ under the two conditions $S = \mu$ and $S = -\mu$, respectively. We still need to select one of them to maximize the lower bound of $\beta$. Since the function $\sqrt{J_0^2(\mu) + J_1^2(\mu)}$ singletonically decreases with respect to $\mu$ for $\mu > 0$, maximizing the similarity is equivalent to taking the smaller value between $\mu_1$ and $\mu_2$. The derivation of this optimal minimum (denoted as $\mu_\text{min}$) is detailed below.

By definition, the condition $|S| = \mu$ implies that one of the two eigenvalues of $\mathbf{W}$ is strictly zero. Given that $\mathbf{D}$ is full-rank, and $\mathbf{P}$ is also full-rank provided the incident angle $\theta$ safisfies $\theta < \frac{\pi}{2}$, this means that $\mathbf{Q}_{\text{BS}} - \frac{1}{R_{\text{sph}}}\mathbf{I}_2$ must be a rank-one matrix.

Apply the eigenvalue decomposition $\mathbf{Q}_{\text{BS}} = \mathbf{V}^T \mathbf{\Lambda}_q \mathbf{V}$, and let $q_1$ and $q_2$ denote the two eigenvalues of $\mathbf{Q}_{\text{BS}}$. To make $\mathbf{Q}_{\text{BS}} - \frac{1}{R_{\text{sph}}}\mathbf{I}_2$ rank-one, first let $R_{\text{sph}}^{-1} = q_1$. In this case, the only non-zero eigenvalue of $\mathbf{W}$ equals its trace:
\begin{equation}
    \lambda(\mathbf{W}) = \frac{\pi d_{\text{ant}}^2}{\lambda} \operatorname{tr}\big( (\mathbf{VPD})(\mathbf{VPD})^T \operatorname{diag}(0, q_2 - q_1) \big).
\end{equation}
Define the matrix $\mathbf{M}$ as  $\mathbf{M} = (\mathbf{VPD})(\mathbf{VPD})^T$, and the above eigenvalue can be directly calculated as $\lambda(\mathbf{W}) = \frac{\pi d_{\text{ant}}^2}{\lambda}(q_2 - q_1)[\mathbf{M}]_{2,2}$. Similarly, when $R_{\text{sph}}^{-1} = q_2$, the corresponding eigenvalue is $\frac{\pi d_{\text{ant}}^2}{\lambda}(q_1 - q_2)[\mathbf{M}]_{1,1}$. With the rank-one condition, $\mu$ satisfies $\mu = \frac{1}{2}|\lambda(\mathbf{W})|$, thus the minimum value $\mu_\text{min}$ is given by:
\begin{equation}
\mu_\text{min} = \frac{\pi d_{\text{ant}}^2}{2\lambda} |q_1 - q_2| \left(\min_{m \in \{1,2\}} [\mathbf{M}]_{m, m}\right).
\end{equation}

Using the derived $\mu_\text{min}$, we can now formulate the scaling factor $t$. Once $S$ is chosen such that $|S|= \mu_\text{min}$, the integral in (\ref{eq:square integral}) depends exclusively on $\mu_\text{min}$. A small $\mu_\text{min}$ indicates minimal phase variation, making the integral nearly uniform across the square; hence, the disk integration contributes $\frac{\pi}{4}$ of the total area. Consequently, $t$ should approach $\frac{4}{\pi}$ as $\mu_\text{min} \to 0$. Conversely, for large $\mu_\text{min}$, the integrand oscillates drastically at the square's corners. By stationary phase method \cite{Balanis16}, the contribution from these corners to the total integral becomes negligible, meaning the disk integral closely approximates the total integral. Therefore, $t \to 1$ as $\mu_\text{min} \to \infty$. Based on these asymptotic behaviors, $t$ can be modeled as a function of $\mu_\text{min}$:
\begin{equation}
    t(\mu_\text{min}) = 1+(\frac{4}{\pi}-1)*e^{-(\mu_\text{min})^2}
\end{equation}

To derive a more tractable criterion, we consider the upper bound of $\mu_\text{min}$, denoted as $\mu^*$. It is evident that $\min_m [\mathbf{M}]_{m,m}$ is upper-bounded by $\frac{1}{2}\text{tr}(\mathbf{M})$. Given the orthogonality of $\mathbf{V}$, the trace can be manipulated as $\text{tr}(\mathbf{M}) = \text{tr}(\mathbf{P}\mathbf{D}\mathbf{D}^T\mathbf{P}^T\mathbf{V}^T\mathbf{V}) = \text{tr}(\mathbf{D}\mathbf{D}^T\mathbf{P}^T\mathbf{P})$. Furthermore, since all entries of $\mathbf{P}$ are inner products of directional vectors, it naturally follows that $\text{tr}(\mathbf{M}) \leq \text{tr}(\mathbf{D}\mathbf{D}^T)$. Consequently, we obtain $\min_m [\mathbf{M}]_{m,m} \leq \frac{N_y^2+N_z^2}{8}$.

In summary, the approximated lower bound of the optimal cosine similarity can be determined as (\ref{eq:cossim expression}).
}

\end{document}